\documentclass[aps,pre,twocolumn,superscriptaddress,showpacs,10pt]{revtex4-1}
\usepackage{graphicx}
\usepackage{epstopdf}
\usepackage{subfigure}
\usepackage{amsmath}
\usepackage{float}
\begin{document}

\preprint{}

\title{Thermodynamic Characterization of Networks Using Graph Polynomials}

\author{Cheng Ye}%
 \email{cy666@york.ac.uk}
\affiliation{
Department of Computer Science, University of York, York, YO10 5GH, UK.
}


\author{C\'{e}sar H. Comin}%
 \email{chcomin@gmail.com}
\affiliation{
Institute of Physics at S$\tilde{a}$o Carlos, University of S$\tilde{a}$o Paulo, PO Box 369, S$\tilde{a}$o Carlos, S$\tilde{a}$o Paulo, 13560-970, Brazil.
}

\author{Thomas K. DM. Peron}%
 \email{thomas.peron@usp.br}
\affiliation{
Institute of Physics at S$\tilde{a}$o Carlos, University of S$\tilde{a}$o Paulo, PO Box 369, S$\tilde{a}$o Carlos, S$\tilde{a}$o Paulo, 13560-970, Brazil.
}

\author{Filipi N. Silva}%
 \email{filipinascimento@gmail.com}
\affiliation{
Institute of Physics at S$\tilde{a}$o Carlos, University of S$\tilde{a}$o Paulo, PO Box 369, S$\tilde{a}$o Carlos, S$\tilde{a}$o Paulo, 13560-970, Brazil.
}

\author{Francisco A. Rodrigues}%
 \email{francisco@icmc.usp.br}
\affiliation{
Institute of Mathematical and Computer Sciences, University of S$\tilde{a}$o Paulo, PO Box 668, S$\tilde{a}$o Carlos, S$\tilde{a}$o Paulo, 13560-970, Brazil.
}

\author{Luciano da F. Costa}%
 \email{ldfcosta@gmail.com}
\affiliation{
Institute of Physics at S$\tilde{a}$o Carlos, University of S$\tilde{a}$o Paulo, PO Box 369, S$\tilde{a}$o Carlos, S$\tilde{a}$o Paulo, 13560-970, Brazil.
}

\author{Andrea Torsello}%
 \email{torsello@dsi.unive.it}
\affiliation{
Department of Environmental Sciences, Informatics and Statistics, Ca' Foscari University of Venice, Dorsoduro 3246 - 30123 Venezia, Italy.
}

\author{Edwin R. Hancock}%
 \email{edwin.hancock@york.ac.uk}
 \thanks{Edwin R. Hancock is supported by a Royal Society Wolfson Research Merit Award.}
\affiliation{
Department of Computer Science, University of York, York, YO10 5GH, UK.
}

\date{September 2015}

\begin{abstract}

In this paper, we present a novel  method for characterizing the evolution of  time-varying complex networks by adopting a thermodynamic representation of  network structure computed from a polynomial (or algebraic) characterization of graph structure. Commencing from a representation of graph structure based on a  characteristic polynomial computed from the normalized Laplacian matrix, we show how the polynomial is  linked to the Boltzmann partition function of a network. This allows us to compute   a number of  thermodynamic quantities for the network, including the average energy and  entropy. Assuming that the system does not change volume, we can also compute the temperature, defined  as the rate of change of entropy with energy. All three thermodynamic variables can be approximated using low-order Taylor series that can be computed using the traces of powers of the Laplacian matrix, avoiding explicit computation of the normalized Laplacian spectrum. These polynomial approximations allow  a smoothed representation of the evolution of networks to be constructed in the thermodynamic space spanned by entropy, energy and temperature. We show how  these thermodynamic variables  can be  computed in terms of simple network characteristics, e.g., the total number of  nodes and node  degree statistics for nodes connected by edges. We apply the resulting  thermodynamic characterization to real-world time-varying networks representing complex systems in the  financial  and biological domains. The study  demonstrates that the method provides an efficient tool for detecting abrupt changes  and characterizing different stages in network evolution.	

\end{abstract}


\maketitle

\section{Introduction}

There has been a vast amount of effort expended on the problems of how to represent networks, and from this representation derive succinct characterizations of network structure and in particular how this structure
evolves with time \cite{Hofstad_2014}\cite{Anan_2009}\cite{Albe_2000}.  Broadly  speaking the  representations and the resulting characterizations are  goal-directed, and have centred around ways of capturing network substructure using clusters, or notions such as hubs and communities \cite{Newman_2003}\cite{Estr_2011}\cite{Feld_1998}\cite{Dehm_2013}. Here the underlying representations are based on the connectivity structure of the network, or statistics that capture the connectivity structure such as degree distributions \cite{Anan_2011}\cite{Anan_2014}.

A more principled approach is to try to characterize the properties of networks using ideas from statistical physics \cite{Kers_1987}\cite{Javarone_2013}. Here the network can be succinctly described using a partition function, and thermodynamic characterizations of the network  such as entropy, total energy and temperature can be derived from the partition function \cite{Miku_2001}\cite{Delv_2011}\cite{Fron_2007}. For example, by interpreting the subgraph centrality as a partition function of a network, the entropy,  internal energy and the Helmholtz free energy are  defined using spectral  graph theory and  various relations between these thermodynamic variables can be obtained \cite{Estr_2007}. However, to embark on this type of analysis, the microstates of the network system must be specified and a clear interpretation of the network thermodynamics provided. This approach has provided some deep insights into network behaviour. For instance, in the work \cite{Bian_2001}, the Bose-Einstein partition function is used to model a Bose gas on a network, and the process of Bose condensation and its quantum mechanical implications have been studied. This model has also been extended to understand processes such as super-symmetry in networks \cite{Bian_2002}.

However, in this context the representation of the network stems from a physical analogy, in which the network provides a Hamiltonian whose eigenstates are occupied according to Bose-Einstein statistics subject to Boltzmann thermalization. Although this type of physical analogy is useful, it does not link directly to the types of representation studied in the graph-theory literature.

\subsection{Related Literature}

Two of the most effective approaches adopted by graph theorists include spectral graph theory and algebraic graph theory \cite{Chun_1997}\cite{Bigg_1993}. These two approaches are intimately related. Both commence from a matrix representation of a graph. In the case of spectral graph theory, it is the eigenvalues and eigenvectors of the matrix that are of interest \cite{Pass_2009}\cite{Brau_2006}.  In algebraic graph theory, a characteristic polynomial is computed from the determinant of the identity matrix minus a multiple of the matrix. The coefficients of this polynomial are determined by symmetric polynomials of the matrix eigenvalues and they provide many useful graph invariants. For example, the coefficients of the Laplacian characteristic polynomial are related to the number of spanning trees and spanning forests in a graph, and particularly, for certain graphs in $(a,b)$-linear classes, the coefficients can be simply expressed in terms of number of nodes in the graph \cite{Oliv_2002}. Spectral methods have been exploited directly and with great effect in complex networks and machine learning. Much of this is due to the close links between graph spectra and random walks on networks. For instance, the heat equation, which governs the behaviour of a continuous time random walk on a network, has been used to model information flow on networks \cite{Esco_2012}. However,  there has been less interest in the algebraic approach. This may be something of an oversight, since there are strong links between algebraic graph theory and number theory, and results from algebraic graph theory can be used to construct important invariants that can be used to  probe network structure. For instance, the Laplacian matrix can be used to construct a zeta function, which can be viewed as an analogue of the Riemann zeta function from number theory \cite{Rose_1997}. This zeta function, is in fact the moment generating function for the heat kernel, and its derivative at origin is linked to the number of spanning trees contained in a network \cite{Xiao_2009}. The Ihara zeta function, which is derived from a characteristic polynomial for the oriented line graph of a network,  can be used to determine the distribution of prime cycles of various  length in a network and is also closely linked to the evolution of  a discrete time quantum walk on a network \cite{Scot_2008}\cite{Ren_2011}\cite{Ren_2011_2}. This latter type of representation has been shown to lift some of the problems in cospectrality of networks encountered if conventional spectral methods are used.

\subsection{Overview}

The aim in this paper is therefore to establish a link between characteristic polynomials from algebraic graph theory, and the thermodynamical analysis of networks. Our characterization commences from  the Boltzmann partition  function $Z(\beta) = tr(\exp\{-\beta \hat{H}\})$ where $\hat{H}$ is the Hamiltonian associated with the graph and $\beta=1/kT$ with $k$ the Boltzmann constant and $T$ the temperature. The Hamiltonian is the  total energy operator, which can be defined in a number of ways. For instance, in quantum mechanics the choice dictated by the Schr\"{o}dinger equation is $\hat{H}=-\nabla^2+U(r,t),$ where $\nabla^2$ is the Laplacian and $U(r,t)$ the potential energy operator.  For a graph, if we specify the node potential energy as the degree matrix, i.e.,  $U(r,t)=D$ and replace the Laplacian by its combinatorial counterpart $L=D-A$, where $A$ is the adjacency matrix, then $\hat{H} = A.$ This choice of Hamiltonian is often used in the H\"{u}ckel molecular orbital (HMO) method \cite{Coul_1978}. An alternative is to assume a graph is immersed in a heat bath with the eigenvalues of its normalized Laplacian matrix as the energy eigenstates. In this case, we set the  potential energy operator $U(r,t)$ to zero, and can identify  $\nabla^2$ with the graph  normalized Laplacian, i.e., $\hat{H} = -\tilde{L}= -D^{-1/2} (D-A) D^{-1/2}$.

With this choice of Hamiltonian and hence partition function, the energy associated with the graph is $E = -\partial \ln Z(\beta)/\partial\beta=-\sum_i p_i \tilde{\lambda_i}$, where $\tilde{\lambda_i}$ denote the eigenvalues of $\tilde{L}$ and $p_i=\exp\{\beta \tilde{\lambda_i}\}/\sum_i \exp\{\beta \tilde{\lambda_i}\}$, i.e., a weighted average of the normalized Laplacian eigenvalues, where the weights associated with the individual eigenvalues are determined by the Boltzmann occupation probabilities. The entropy is given by $S=k\{\ln Z(\beta) +\beta E\}$.

We characterize the graph using the Ihara zeta function $R(\beta) = \det(I-\beta\tilde{L})$. We show in our analysis that $Z(\beta) \simeq -\ln R(\beta)+N$, where $N$ is the graph size and as a result both  the energy and entropy can be expanded as  power series in $\beta$. The leading coefficients of the two series are determined by the sum of the reciprocal of the degree-products for nodes  forming edges and triangles in the  graph. The coefficients of the increasing powers of $\beta$ depend on the frequencies  increasingly large substructures. The higher the degrees of the nodes forming these structures, the smaller the associated weight. Hence high degree structures are energetically more favourable than low degree ones (because they have lower reciprocal of the degree product). Also larger structures are also energetically more favourable.

The expressions  derived for energy and entropy of the network depend only on the assumed model for Hamiltonian of the system, and the approximations needed to express the partition function in terms of the characteristic polynomial associated with the normalized Laplacian  of the graph. Hence the energy and entropy can be used as a characterization of  structure for any set of networks.  However, in our experiments  we study the time-evolution of networks with  fixed numbers of nodes.  This is not an entirely uncommon situation, and arises where networks are used to abstract systems with a known set of states or components. In the financial network example, the nodes are stock traded over a 6000-day  period, and in the second example the nodes represent genes expressed by fruit flies at different stages in their development. In this set-up we require a natural way of measuring fluctuations in network structure with time.

For a thermodynamic system with freedom to vary its volume, temperature and pressure, the change in internal energy is given by $dE=TdS-\mathcal{P}d\mathcal{V}+mdN$ where $T$ is the temperature, $\mathcal{P}$ the pressure, $d\mathcal{V}$ the change in volume, $m$ the particle mass and $dN$ the change in the number of particles.  When the number of particles  and volume are  fixed, we have an isochoric process, and the temperature is the rate of change of energy with entropy. With the expressions for these two quantities derived from the partition function, the isochoric temperature is  also determined  by a simple expression involving the frequencies of edges and triangles  of different degree configuration. One way to picture this system is a thermal distribution across the energy states corresponding to the normalized Laplacian eigenvalues.  Large  changes in temperature are hence associated with a) large changes in the number of triangles compared to the number of edges, and b) when  the  average degree of  the nodes changes significantly. Hence the temperature fluctuation between graphs in a sequence   is sensitive to changes in internal structure of the network. We show that our method in fact smooths the time dependance of the thermodynamic characterization, so we present the global thermodynamic analysis in a computationally efficient and tractable way.

So, to summarize we present a method motivated by thermodynamics for characterizing time sequences of networks. Although it is not a model of network evolution, it may provide the building blocks for such a model.
The approach has some similarities to that reported by Javarone and Armano \cite{Javarone_2013} who use the classical limits of quantum models of gases as analogues to analyze complex networks. However, rather than using the classical Boltzmann distribution and the normalized Laplacian characteristic polynomial as the basis of their model, they base their model on a fermionic system. Finally, we note that the notion of temperature used in our work is not the physical temperature of the system, but a means of gauging fluctuations in network structure with time.

The remainder of the paper is structured as follows. In Sec. II, we first show how the Boltzmann partition function is linked to the characteristic polynomial of the normalized Laplacian matrix of graphs. With this to hand, we then provide a detailed account of the  development of a number of thermodynamic variables of networks, i.e., the average energy, thermodynamic entropy and temperature. In Sec. III, we apply the resulting thermodynamic characterization to  a number of real-world time-varying networks, including the New York Stock Exchange (NYSE) data and  the fruit fly life cycle gene expression data. Finally, in Sec. IV, we conclude the paper  and  make suggestions for future work.

\section{Thermodynamic Variables of Complex Networks}

In this section, we provide a detailed development of  how we compute thermodynamic quantities for a network,  including the thermodynamic entropy, average energy and temperature, commencing from a characteristic polynomial representation of network structure.  First, we provide some preliminaries on how  graphs can be  represented using the normalized Laplacian matrix. We then explain how the Boltzmann partition function can be used to describe the thermalization of the population of the energy  microstates of network as represented by its Hamiltonian. The key step in establishing our  thermodynamic characterization of network evolution, is to show a relationship between the partition function and the characteristic polynomial for the network. Normally, the thermalization process arises via the analogy of emersing the network in heat bath, with the adjacency matrix eigenvalues playing the role of energy eigenstates and the  thermal population of the energy levels being controlled by the Boltzmann distribution. Here we aim to make a connection between the heat bath analogy and an alternative graph representation based on a characteristic polynomial. This is a powerful approach since there are several alternative matrix representations of graphs, and their characteristic polynomials together with the closely related zeta-function representations have been extensively studied in graph theory \cite{Scot_2008}\cite{Ren_2011}\cite{Ren_2011_2}. Our approach therefore allows these potentially rich representations to be investigated from the thermodynamic perspective. Specifically, we show how the partition function can be approximated by the characteristic polynomial associated with the  normalized Laplacian matrix for the network. This picture of the heat bath emerges when the Hamiltonian is  the negative normalized Laplacian. From this starting point and using  the network partition function approximation, we derive the expressions for the network average energy and entropy, and under the assumption of constant volume determine the network  temperature by measuring fluctuations in entropy and average energy. We show for networks of approximately  constant size, each of these thermodynamic quantities can be   computed using  simple network statistics, including the number of nodes and   node degree statistics.

\subsection{Initial Considerations}

Let $G(V,W)$ be an undirected graph with node set $V$ and edge set $W\subseteq V \times V$, and $N=|V|$  is the total number of nodes. The adjacency matrix $A$ of graph $G$ is defined as
\begin{equation}
A_{uv}=\left\{ \begin{array}{cl}
 1 &\mbox{ if $(u,v)\in W$ } \\
 0 &\mbox{ otherwise.}
       \end{array} \right.
\end{equation} The degree of node $u$ is $d_u=\sum_{v \in V}A_{vu}.$

Then, the normalized Laplacian matrix $\tilde{L}$ is defined as $\tilde{L}=D^{-1/2}LD^{-1/2}$ where $L=D-A$ is the Laplacian matrix and $D$ denotes the degree diagonal matrix whose elements are given by $D(u,u)=d_u$ and zeros elsewhere. The elementwise expression of $\tilde{L}$ is
\begin{equation}\label{normalized laplacian}
\tilde{L}_{uv}=\left\{ \begin{array}{cl}
 1                        &\mbox{ if $u=v$ and $d_v\neq0$ } \\
 -\frac{1}{\sqrt{d_ud_v}} &\mbox{ if $u\neq v$ and $(u,v)\in W$} \\
 0                        &\mbox{ otherwise.}
       \end{array} \right.
\end{equation} The normalized Laplacian matrix $\tilde{L}$ and its spectrum yield a number of very useful graph invariants for a finite graph.  For example, the eigenvalues for the  graph normalized  Laplacian  are real numbers,  bounded between 0 and 2. Moreover, the multiplicity of zero eigenvalue of $\tilde{L}$ is the number of connected components in a graph $G$ while the multiplicity of eigenvalue equal to 2 is the bipartite connected component number in $G$ ($G$ has at least two nodes) \cite{Chun_1997}.

\subsection{Boltzmann Partition Function}

In statistical mechanics, the canonical partition function associated with the Boltzmann factor of a system is
\begin{equation}
   Z=\sum_i e^{-\beta E_i}
\end{equation} where $\beta=1/kT$ is proportional to  the reciprocal of the temperature $T$ with $k$ the Boltzmann constant,  and $E_i$ denotes the total energy of the system when it is in microstate $i$. Moreover,  the partition function can be formalized as a trace over the state space:
\begin{equation}
Z(\beta) = tr(\exp\{-\beta \hat{H}\})
\end{equation} where $\hat{H}$ is the Hamiltonian operator and $\exp\{\cdot\}$ represents the matrix exponential.

The Hamiltonian operator of a graph may be defined in a number of ways. In quantum mechanics, one choice dictated by the Schr\"{o}dinger equation is $$\hat{H}=-\nabla^2+U(r,t).$$ If we set the  potential energy operator $U(r,t)$ to zero,  we can identify  $\nabla^2$ with the graph Laplacian in either  its combinatorial or normalized form. With this choice we  obtain
$$\hat{H} = -L$$ or
\begin{equation}\label{hamiltonian}
\hat{H} = -\tilde{L}.
\end{equation} Alternatively, we can specify the node potential energy operator as  the degree matrix, i.e.,  $U(r,t)=D$, with the result that $$\hat{H} = A.$$ This choice of Hamiltonian is often used in  H\"{u}ckel molecular orbital (HMO) method \cite{Coul_1978}. Generally, in this case $\hat{H} = c_1 I + c_2 \mathcal{A}$ where  $\mathcal{A}$ is the adjacency matrix of a graph representing the carbon skeleton of the molecule and $c_1,c_2$ are constants.

In our analysis we let the Hamiltonian operator $\hat{H} = -\tilde{L}$ as in Eq.(\ref{hamiltonian}), as a result, the Boltzmann partition function takes the form
\begin{equation}\label{partition function}
  Z(\beta) = tr(\exp\{\beta \tilde{L}\}).
\end{equation} Although most of the aggregate thermodynamic variables of the system, such as the average energy and entropy, can be expressed in terms of the partition function or its derivatives, deriving expressions for these variables directly from Eq.(\ref{partition function}) can be  computationally difficult. A more convenient route is to adopt an alternative graph representation based on a characteristic
polynomial. In this way we approximate the Boltzmann partition function, so that the computation for thermodynamic variables can be simplified.

It is important to stress that making use of the statistical mechanical analysis usually requires a specification of the microscopic configurations of a thermodynamic system together with a clear physical interpretation of their meaning. In this paper, we do not dwell on the microstates of the  thermodynamic system  arise or how they are populated.  Briefly, our Hamiltonian is the negative of the normalized Laplacian, and one physical interpretation of our model would be of a graph immersed in a heat bath with the normalized Laplacian eigenvalues as energy eigenstates. The graph is subject to thermalization via the Boltzmann distribution. Our main concern is though to understand how to approximate the partition function of the resulting system so as to render thermodynamic analysis tractable. Although we do define a Hamiltonian for the system, our basic representation of the graph is in terms of the  characteristic polynomial. We show how the characteristic polynomial can be used to approximate the Boltzmann partition function when the graph is immersed in a heat bath. Here the polynomial coefficients are themselves symmetric polynomials of the normalized Laplacian eigenvalues, and the polynomial variable is linked to the temperature of the heat bath. As we will show in our experiments, this approximation effectively smooths the time dependance of the network evolution, by allowing the thermodynamic variables to be approximated by low-order polynomials.

\subsection{Characteristic Polynomial of Normalized Laplacian Matrix}

The characteristic polynomial of the normalized Laplacian matrix $\tilde{L}$ of a graph, denoted by $P_{ch}(x)$, is the polynomial defined by
\begin{equation}
  P_{ch}(x) = \det(xI-\tilde{L})
\end{equation} where $I$ indicates the identity matrix and $x$ is the polynomial variable.

At this point it is worth noting that polynomial characterizations are also central to the definition of various types of zeta function of a graph. For instance, the determinant expression for the reciprocal of the Ihara zeta functions of a graph $G$  \cite{Ren_2011} is
\begin{equation}
  \zeta^{-1}(x)=\det(I-xB)
\end{equation} where $B$ is the Hashimoto's edge adjacency operator on the oriented line graph of $G$. By replacing the Hashimoto operator with the normalized Laplacian operator $B=\tilde{L}$, we immediately obtain
\begin{equation}
  \zeta^{-1}(x)=\det(I-x\tilde{L}).
\end{equation} Therefore the characteristic polynomial of the normalized Laplacian matrix  and the above zeta function of graph $G$ are related by
\begin{equation*}
  P_{ch}(x) = x^N \det(I-\frac{1}{x}\tilde{L}) = x^N \zeta^{-1}(\frac{1}{x})
\end{equation*} where $N$ is the number of nodes in graph $G$.

Here we use $R(x)$ to denote the Ihara-zeta-function determinant $\det(I-\frac{1}{x}\tilde{L})$ and refer to it as the quasi characteristic polynomial of the normalized Laplacian matrix. To show that $R(x)$ can be employed as an efficient tool for approximating the Boltzmann partition function in Eq.(\ref{partition function}), we first note that for a square matrix $M$, the determinant can be calculated by
$$\det(M)=\exp\{tr(\ln M)\}.$$ Thus, we have
\begin{equation}
R(x) = \exp\{tr[\ln(I-\frac{1}{x}\tilde{L})]\}.
\end{equation} Recalling   the classical Mercator series for the matrix  logarithm  of  $I+M$ $$\ln(I+M)=M-\frac{M^2}{2}+\frac{M^3}{3}-\cdots, \quad \rho(M) < 1,$$ where $\rho(M)$ indicates the spectral radius of $M$, which is equal to the largest absolute value of the eigenvalues of $M$. Since the normalized Laplacian matrix has eigenvalues between 0 and 2 \cite{Chun_1997}, the matrix Mercator series holds if and only if $\rho(\frac{1}{x}\tilde{L})<1$, i.e., $|\frac{1}{x}|<\frac{1}{2}$.

To develop these ideas  one step further, if we let $\frac{1}{x}=\beta$, the quasi characteristic polynomial of the normalized Laplacian matrix can then be expressed as
\begin{equation}\label{modified characteristic polynomial}
R(\beta) = \exp\{tr(-\beta\tilde{L}-\frac{1}{2}\beta^2\tilde{L}^2-\frac{1}{3}\beta^3\tilde{L}^3-\cdots)\}.
\end{equation} Moreover, using  the first-order MacLaurin formula to expand the matrix exponential, i.e., $$\exp M=I+M+\frac{M^2}{2!}+\frac{M^3}{3!}+\cdots$$ where $M$ is an arbitrary square matrix, we can  immediately re-write the Boltzmann partition function Eq.(\ref{partition function}) in the following way:
\begin{equation}\label{revised partition function}
Z(\beta) = tr(I+\beta\tilde{L}+\frac{1}{2!}\beta^2\tilde{L}^2+\frac{1}{3!}\beta^3\tilde{L}^3+\cdots).
\end{equation}

By comparing the expressions  in Eq.(\ref{modified characteristic polynomial}) and Eq.(\ref{revised partition function}), the Boltzmann partition function can then be calculated from the quasi characteristic polynomial of the normalized Laplacian matrix as follows:
\begin{eqnarray}
Z(\beta) &=& tr(I)+ tr(\beta\tilde{L}+\frac{1}{2!}\beta^2\tilde{L}^2+\cdots) \nonumber\\
             &=& N-\ln R(\beta) +r(\beta),
\end{eqnarray} where $r(\beta)$ denotes the residual. More explicitly, the residual is computed by
\begin{eqnarray*}
  r(\beta) &=& \sum_{n=3}^\infty(\frac{1}{n!}-\frac{1}{n})\beta^ntr(\tilde{L}^n) \nonumber \\
           &=& - \sum_{n=3}^\infty \frac{\beta^n}{n}\biggl[1-\frac{1}{(n-1)!}\biggl]tr(\tilde{L}^n) \nonumber \\
           &=& -\frac{\beta^3}{6}tr(\tilde{L})-\frac{5\beta^4}{24}tr(\tilde{L}^2)-\cdots.
\end{eqnarray*} As a result, when $|\beta|$ takes on a small value,  we have $$\lim_{\beta\rightarrow 0} \frac{r(\beta)}{\ln R(\beta)} = 0,$$ i.e., $r(\beta)=o[\ln R(\beta)]$. This implies that the partition function is approximately equal  to the negative of natural logarithm of the quasi characteristic polynomial plus a constant:
\begin{equation}\label{approximation}
  Z(\beta) \simeq -\ln R(\beta)+N.
\end{equation}

To conclude this subsection, it is worth discussing the validity of the above approximation. We have shown that the requirements a) $|\beta|<\frac{1}{2}$ and b) $r(\beta)=o[\ln R(\beta)]$ are essential to making this approximation valid, which implies that the value of $\beta$ must be small. In Sec. III we will provide an empirical analysis showing that this condition is well satisfied for a number of real-world complex networks.

\subsection{Thermodynamic Variables of Complex Networks}

For thermodynamics, a thermodynamic state of a system can be fully described by an appropriate set of principal parameters known as thermodynamic variables. These include the average energy, entropy and temperature. In this subsection, we give a detailed development showing how these thermodynamic state variables are derived from the approximate partition function and how they can be computed via simple network statistics.

To commence, we recall that given a partition function $Z(\beta)$, the average energy $E$ of a system $G$ is obtained by taking the partial derivative of the logarithm of the partition function   with respect to $\beta$, i.e.,
\begin{equation}\label{original average energy}
  E(G) = -\frac{\partial \ln Z(\beta)}{\partial\beta}.
\end{equation} Moreover, the thermodynamic entropy $S$ is obtained by
\begin{equation}\label{original entropy}
  S(G) = k\{\ln Z(\beta) +\beta E(G)\}
\end{equation} where $k$ denotes the Boltzmann constant.

\subsubsection{Temperature}

The thermodynamic temperature $T$, measures fluctuations in network structure with time. More specifically, suppose that $G_1$ and $G_2$ represent the structure of a time-varying system at two consecutive epochs
$t_1$ and $t_2$ respectively. For a thermodynamic system of constant number of particles, we recall the fundamental thermodynamic relation $dE=TdS-\mathcal{P}d\mathcal{V}$, where $\mathcal{P}$ and $\mathcal{V}$ denote the pressure and volume respectively. The volume is a concept generally considered in the context of ideal gases and many thermodynamic processes could result in a change in volume. Here we consider the network under study $G$ as a closed system and from $G_1$ to $G_2$ it undergoes a constant-volume process (isochoric process) during which the system volume remains constant.

It is important to stress that this equation holds and is  valid for both reversible and irreversible processes for a closed system, since $E$, $T$, $S$, $\mathcal{P}$ and $\mathcal{V}$ are all state functions and are independent of thermodynamic path.  As a result,  for the path from $G_1$ to $G_2$ we have $d\mathcal{V} = 0$ and  $dE=TdS$. For example, when an ideal gas undergoes an isochoric process, and the quantity of gas remains constant, then the energy increment is proportional to the increase in temperature and pressure. As a result, the  reciprocal of the  temperature $T$ is the  rate of change of entropy with   average energy,
subject to  the condition that the volume and number of  particles are  held constant, i.e.,
\begin{equation}\label{original temperature}
\frac{1}{T(G_1,G_2)}=\frac{dS}{dE}=\frac{S_1-S_2}{E_1-E_2}.
\end{equation} This definition can be applied to evolving complex networks which do not change significantly in size during their evolution.

To further develop the temperature expression, we first compute the change in entropy
\begin{eqnarray}
  S_1-S_2 &=& k\{\ln Z_1(\beta) +\beta E_1(G)\}-k\{\ln Z_2(\beta) +\beta E_2(G)\} \nonumber\\
          &=& k\{\ln\frac{Z_1}{Z_2}+\beta(E_1-E_2)\}.
\end{eqnarray} Note, that in our development, the partition function is approximated by $Z(\beta) \simeq -\ln R(\beta)+N$. Therefore, we have
\begin{eqnarray*}
  \ln\frac{Z_1}{Z_2} &\simeq& \ln\frac{N-\ln R_1}{N-\ln R_2} \nonumber\\
                          &=& \ln N + \ln(1-\frac{1}{N}\ln R_1)-\ln N- \ln(1-\frac{1}{N}\ln R_2)\nonumber\\
                          &=& \ln(1-\frac{1}{N}\ln R_1)- \ln(1-\frac{1}{N}\ln R_2).
\end{eqnarray*} The term $\frac{1}{N}\ln R$ is close to zero since we assume that $|\beta|$ is small. As a result,  using the Mercator series, we obtain $\ln(1-\frac{1}{N}\ln R)\simeq-\frac{1}{N}\ln R,$ leading to the result that
\begin{eqnarray}
  \ln\frac{Z_1}{Z_2} &\simeq& -\frac{1}{N}\ln R_1 + \frac{1}{N}\ln R_2 \nonumber\\
                     &=&  \frac{1}{N}\ln\frac{R_2}{R_1} =  \frac{1}{N}\ln(1+\frac{R_2-R_1}{R_1}) \nonumber\\
                     &\simeq& \frac{1}{N}\cdot\frac{R_2-R_1}{R_1}
\end{eqnarray} where $R_2 - R_1$ is the difference between the values for the quasi characteristic polynomial $R(\beta)$ at times $t_1$ and $t_2$.

Next, we calculate the energy
\begin{eqnarray}
  E(G) &\simeq& -\frac{\partial \ln (N-\ln R)}{\partial\beta} \nonumber\\
       &=& - \frac{1}{N-\ln R}\cdot\frac{\partial(N-\ln R)}{\partial\beta}\nonumber\\
       &=& \frac{1}{N-\ln R}\cdot\frac{\partial\ln R}{\partial\beta} \nonumber\\
       &=& -\frac{1}{N-\ln R}\cdot\sum_{n=1}^\infty\beta^{n-1}tr(\tilde{L}^n).
\end{eqnarray} Since the value for $\beta$ is always small,   then $\ln R(\beta)\ll N$, and as a result the average energy expression is
\begin{equation}\label{simplified energy}
  E(G) = -\frac{1}{N}\sum_{n=1}^\infty\beta^{n-1}tr(\tilde{L}^n).
\end{equation} As a result, the difference between network energy $E$ at times $t_1$ and $t_2$, is
\begin{equation}
E(G_1)-E(G_2)= E_1-E_2 = -\frac{1}{N}[P_1(\beta)-P_2(\beta)]
\end{equation} where $P(\beta)=\sum_{n=1}^\infty\beta^{n-1}tr(\tilde{L}^n).$

Then, we compute the temperature using Eq.(\ref{original temperature}), with the result that
\begin{eqnarray}
  \frac{1}{T(G_1,G_2)} &=& \frac{k\{\ln\frac{Z_1}{Z_2}+\beta(E_1-E_2)\}}{E_1-E_2} \nonumber\\
                       &\simeq&  k\beta - k \cdot \frac{\frac{R_2}{R_1}-1}{P_1-P_2}.
\end{eqnarray} Both the quasi characteristic polynomial $R(\beta)$ and the polynomial $P(\beta)$ can be expanded as power series, expressed as  sums of  traces of the powers of the normalized Laplacian matrix of the network. Expanding the two polynomials to third order requires the following traces:
\begin{eqnarray}
  tr(\tilde{L})&=& N, \nonumber\\
  tr(\tilde{L}^2)&=& N+J, \nonumber\\
  tr(\tilde{L}^3)&=& N+3J-Q
\end{eqnarray} where $$J=\sum_{u,v}\frac{A_{uv}}{d_ud_v}$$ and $$Q=\sum_{u,v,w}\frac{A_{uv}A_{vw}A_{wu}}{d_ud_vd_w}$$ respectively \cite{Han_2012}\cite{Ye_2014}. Expanding  $R(\beta)$ to third order, we find
\begin{widetext}
\begin{eqnarray}
  \frac{R_2}{R_1} &=& \frac{\exp\{tr(-\beta\tilde{L_2}-\frac{\beta^2}{2}\tilde{L_2}^2-
  \frac{\beta^3}{3}\tilde{L_2}^3)\}}{\exp\{tr(-\beta\tilde{L_1}-\frac{\beta^2}{2}\tilde{L_1}^2-\frac{\beta^3}{3}\tilde{L_1}^3)\}} \nonumber\\
                  &=& \exp\biggl\{\beta[tr(\tilde{L_1})-tr(\tilde{L_2})]+\frac{\beta^2}{2}[tr(\tilde{L_1}^2)-tr(\tilde{L_2}^2)]
                  +\frac{\beta^3}{3}[tr(\tilde{L_1}^3)-tr(\tilde{L_2}^3)]\biggl\} \nonumber\\
                  &=& \exp\biggl\{\frac{\beta^2}{2}(J_1-J_2)+\frac{\beta^3}{3}[3(J_1-J_2)-(Q_1-Q_2)]\biggl\}.
\end{eqnarray}
\end{widetext} Similarly, for $P(\beta)$ we obtain
\begin{equation}
P_1-P_2 = \beta(J_1-J_2)+\beta^2[3(J_1-J_2)-(Q_1-Q_2)].
\end{equation} As a result, the reciprocal of the temperature is given by
\begin{widetext}
\begin{equation}
  \frac{1}{T(G_1,G_2)} = k\beta+k\cdot\frac{1-\exp\biggl\{\frac{\beta^2}{2}(J_1-J_2)+\frac{\beta^3}{3}[3(J_1-J_2)-(Q_1-Q_2)]\biggl\}}{\beta(J_1-J_2)+\beta^2[3(J_1-J_2)-(Q_1-Q_2)]}.
\end{equation}
\end{widetext} Since $T=1/k\beta$, the second term on the right-hand side must vanish. As a consequence, we have that
\begin{equation}
\frac{\beta^2}{2}(J_1-J_2)+\frac{\beta^3}{3}[3(J_1-J_2)-(Q_1-Q_2)] = 0.
\end{equation} Firstly, when $J_1-J_2=Q_1-Q_2=0$, i.e., graphs $G_1$ and $G_2$ are identical, $T=1/k\beta$  holds. In other words, there are no structural differences between graphs $G_1$ and $G_2$. A second trivial solution is obtained by $\beta = 0$, implying that the temperature $T=1/k\beta$ goes to infinity. Finally, the nontrivial solution is
\begin{equation}
\beta=-\frac{3(J_1-J_2)}{6(J_1-J_2)-2(Q_1-Q_2)},
\end{equation} which leads to the following expression for the temperature
\begin{equation}\label{temperature}
  T(G_1,G_2)=\frac{1}{k\beta}=-\frac{2}{k}+\frac{2}{3k}\cdot\frac{Q_1-Q_2}{J_1-J_2}.
\end{equation} Here $J_1-J_2$ and $Q_1-Q_2$ represent the change in quantities $J$ and $Q$ when graph $G_1$ evolves to $G_2$ respectively:
\begin{widetext}
\begin{eqnarray}
  J_1-J_2 &=& \sum_{u_1,v_1\in V_1}\frac{A_{u_1v_1}}{d_{u_1}d_{v_1}} - \sum_{u_2,v_2\in V_2}\frac{A_{u_2v_2}}{d_{u_2}d_{v_2}} \\
  Q_1-Q_2 &=& \sum_{u_1,v_1,w_1\in V_1}\frac{A_{u_1v_1}A_{v_1w_1}A_{w_1u_1}}{d_{u_1}d_{v_1}d_{w_1}} - \sum_{u_2,v_2,w_2\in V_2}\frac{A_{u_2v_2}A_{v_2w_2}A_{w_2u_2}}{d_{u_2}d_{v_2}d_{w_2}}.
\end{eqnarray}
\end{widetext}

The temperature measures fluctuations in the internal structure of the time-evolving  network, and depends on the ratio of total change of degree statistics for nodes that form triangles and for nodes connected by  edges in the network.  This is a direct consequence of the fact that we have truncated our series expansion of the partition function  with third order.  If we had continued the expansion to higher order, then the temperature would reflect this and contain terms in the numerator and denominator corresponding to changes in the number of cliques of size larger than 3. By adjusting temperature in this way, we take account of fluctuations from the expected value of temperature $T=1/k\beta$. When combined with the polynomial approach, this has the effect of smoothing the time dependance of the thermodynamic representation.

\subsubsection{Energy and Entropy}

Finally, in order to calculate the  network average energy, we substitute the obtained $\beta$ into Eq.(\ref{simplified energy}) and again remove the terms that have powers larger than 3, with the result that
\begin{equation}\label{average energy}
  E(G) = -\frac{1}{N}[N+\beta(N+J)+\beta^2(N+3J-Q)].
\end{equation} Similarly, for the thermodynamic entropy, we have
\begin{eqnarray*}
  S(G) &=&  k\{\ln Z(\beta) +\beta E(G)\}      \nonumber\\
       &\simeq& k\{\ln (N-\ln R)+\beta E\}   \nonumber\\
       &\simeq&  k\biggl\{\ln N-\frac{1}{N}\ln R + \beta E\biggl\} \nonumber\\
       &=&  k\biggl\{\ln N-\frac{1}{N}\sum_{n=1}^\infty(1-\frac{1}{n})\beta^n tr(\tilde{L}^n)\biggl\},
\end{eqnarray*} and expanding to third order,
\begin{equation}\label{entropy}
  S(G) = k\ln N-\frac{k}{N}\biggl[\frac{\beta^2}{2}(N+J)+\frac{2\beta^3}{3}(N+3J-Q)\biggl].
\end{equation}

In order to obtain a better understanding of these network thermodynamic measures, it is interesting to explore how the average energy and entropy are bounded for graphs of a particular size, and in particular which topologies give the maximum and minimum values of the  energy and entropy  (we consider connected graphs only).

From Eq.(\ref{average energy}) and Eq.(\ref{entropy}), when the quantity $J$ is minimal and quantity $Q$ reaches its maximal value,   both the energy and the  entropy reach their maximum values. This occurs when each pair of graph nodes is connected by an edge,  and this means that  the graph is complete. On the other hand, when $J$ and $Q$ respectively take on their maximal and minimal values, the energy and entropy reach their minimum values. This occurs when the structure is a string.

The  maximum  and minimum average energies and entropies corresponding to these cases are as follows. For a complete graph $K_n$, in which each node has degree $n-1$,  we have that
$$E(K_n)=-\biggl[1+\frac{n}{n-1}\beta+\frac{n^2}{(n-1)^2}\beta^2\biggl]$$ and $$S(K_n)=k\ln n-k\biggl[\frac{n}{2(n-1)} \beta^2+\frac{2n^2 }{3(n-1)^2}\beta^3\biggl].$$ Turning our attention to the case of a string $P_n$ ($n\geq2$), in which two terminal nodes have degree 1 while the remainder  have degree 2, we have that
$$E(P_n)=-\biggl[1+\frac{3n+1}{2n}\beta+\frac{5n+3}{2n}\beta^2\biggl]$$ and $$S(P_n)=k\ln n-k\biggl[\frac{(3n+1)}{4n}\beta^2+\frac{(5n+3)}{3n}\beta^3\biggl].$$ As a result, the average energy and entropy of  graphs with $N$ nodes are bounded as follows:
\begin{widetext}
\begin{eqnarray}
  -\biggl[1+\frac{3N+1}{2N}\beta+\frac{5N+3}{2N}\beta^2\biggl] &\leq&    E(G) \leq  -\biggl[1+\frac{N}{N-1}\beta+\frac{N^2}{(N-1)^2}\beta^2\biggl] \\
  k\ln N-k\biggl[\frac{(3N+1)}{4N}\beta^2+\frac{(5N+3)}{3N}\beta^3\biggl] &\leq&    S(G) \leq k\ln N-k\biggl[\frac{N}{2(N-1)} \beta^2+\frac{2N^2 }{3(N-1)^2}\beta^3\biggl]
\end{eqnarray}
\end{widetext} where the lower bounds are achieved by strings, while the upper bounds are obtained for complete graphs.

There are a number of points to note concerning the development above. One of the most fundamental aspects of the presented thermodynamic measurements is the interplay between quantities $J$ and $Q$. The first represents the direct   connections  of nodes (also known as generalized Randi\'{c} indices \cite{Cave_2010}), while the second  is related to the number of triangles. Both measurements are weighted by their joint degrees.

To provide a deeper  intuition concerning the physical  meaning of our thermodynamic analysis in terms of changes in graph structure, we provide some examples. We  commence by considering  a regular graph with $N$ nodes in which each  node has  the same degree $m$ ($N \cdot m$ must be an even number). In this case, the quantity $J$ is  the sum of existing edges weighted by the network average degree $m$: $$J_{reg}=\frac{N}{m}.$$ This result holds for both trees and cyclic $n$-dimensional lattices. On the other hand, the calculation of $Q$ depends on the nature of the connections for the regular networks. For lattices connecting nodes at distance $d=1$ (first neighborhood) and for all trees, $Q_{reg}$ = 0 (since there are no triangles). For other regular networks the  value of $Q$  depends on the number of triangles in the network $N_{tri}$, i.e., $$Q_{reg}=\frac{6N_{tri}}{m^3}.$$ The multiplicative factor 6 is needed as the summation in the equation of  $Q$ considers each edge $(u,v)$ two times, also because the summation is taken over all edges, and each triangle is counted 3 times. Moreover, when the regular network is a lattice of neighborhood distance $d\geq 2$, $$Q_{reg}(d)=\frac{2NN_{tri}(d)}{m^3},$$ where $N_{tri}(d)$ is the number of triangles of each repeated element. Finally, for the cyclic 1D-lattice with connection distance $d$, the number of triangles each node participates is given by $N_{tri}(d)=3(d-1)[(d-1)+1]/2=3d(d-1)/2$, the average degree is $m=2d$, thus the quantity $Q$ is evaluated as follows: $$Q_{lat-1D}(d)=\frac{3(d-1)}{8d^3}.$$

As noted earlier, this analysis is based on a power series  expansion of the partition function up to order 3. Clearly, to develop a realistic thermodynamic model for structures in which triangles are absent by reason of construction, then the expansion should be taken to higher order. Unfortunately, this renders analysis of the traces appearing in the partition function in terms of degree statistics intractable \cite{Han_2012}\cite{Ye_2014}. An alternative would be to use the Ihara zeta function \cite{Scot_2008} as a network characterization. Here the underlying characteristic is computed from the adjacency matrix of the oriented line graph for a network. The polynomial coefficients are related to the numbers of prime cycles of varying length in a network \cite{Ren_2011}.

To summarize, in this section we have proposed a novel method for characterizing the evolution of complex networks by employing thermodynamic variables. Specifically, we commence from a quasi characteristic polynomial of the normalized Laplacian matrix of a network and  show this polynomial can be used as a tool for approximating the Boltzmann   partition function on the network, when we identify Hamiltonian operator with the normalized Laplacian operator. Then, using the approximate network partition function, we develop the expressions for the network average energy and entropy.  The thermodynamic temperature measures fluctuations via the changes in the connectivity pattern of the network, and is determined by the distribution of  node degree. We show  that these thermodynamic variables are expressed in terms of simple network features, including the number of nodes and the degree statistics for connected  nodes.

\section{Experiments and Evaluations}

We have derived  expressions for the thermodynamic entropy, average energy and temperature of time-evolving complex networks. In this section, we explore whether the resulting  characterization can be employed to provide a useful  tool for better understanding the evolution of dynamic networks. Specifically, we aim at applying the novel thermodynamic method to a number of real-world time-evolving networks in order to analyze whether abrupt changes in structure  or  different stages in  network evolution can be efficiently characterized. In this section, to simplify the calculation, we let the Boltzmann constant $k=1$.

\subsection{Datasets}

We commence by giving a brief  overview of the datasets used for experiments here. We use two different datasets, both are extracted from real-world complex systems.

\begin{figure}[h]
\centering
 \includegraphics[scale=.5]{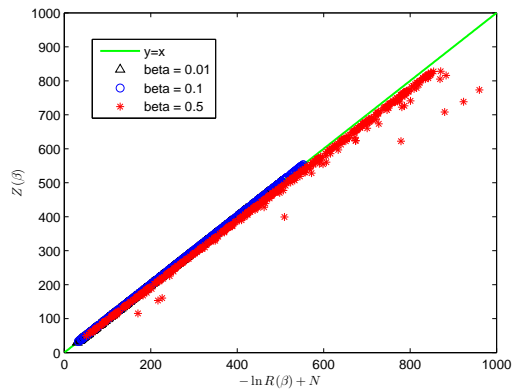}
 \caption{(Color online) The scatter plot of Boltzmann partition function associated with normalized Laplacian operator in Eq.(\ref{partition function}) and the normalized Laplacian quasi characteristic polynomial given by Eq.(\ref{modified characteristic polynomial}) for different $\beta$ for Erd\H{o}s-R\'{e}nyi and Barab\'{a}si-Albert random graphs. Black triangles: $\beta = 0.01$; blue circles: $\beta = 0.1$; red stars: $\beta = 0.5.$}
\end{figure}

{\bf Dataset 1:} Is extracted from a  database consisting of the daily prices of 3799 stocks traded on the  New York Stock Exchange (NYSE). This data has been well analyzed in Ref.\cite{Silva_2015}, which has provided an empirical investigation studying the role of communities in the structure of the inferred NYSE stock market. The authors have also defined a community-based model to represent the topological variations of the market during financial crises.

\begin{figure}[h]
\centering
 \subfigure[]{\includegraphics[scale=.45]{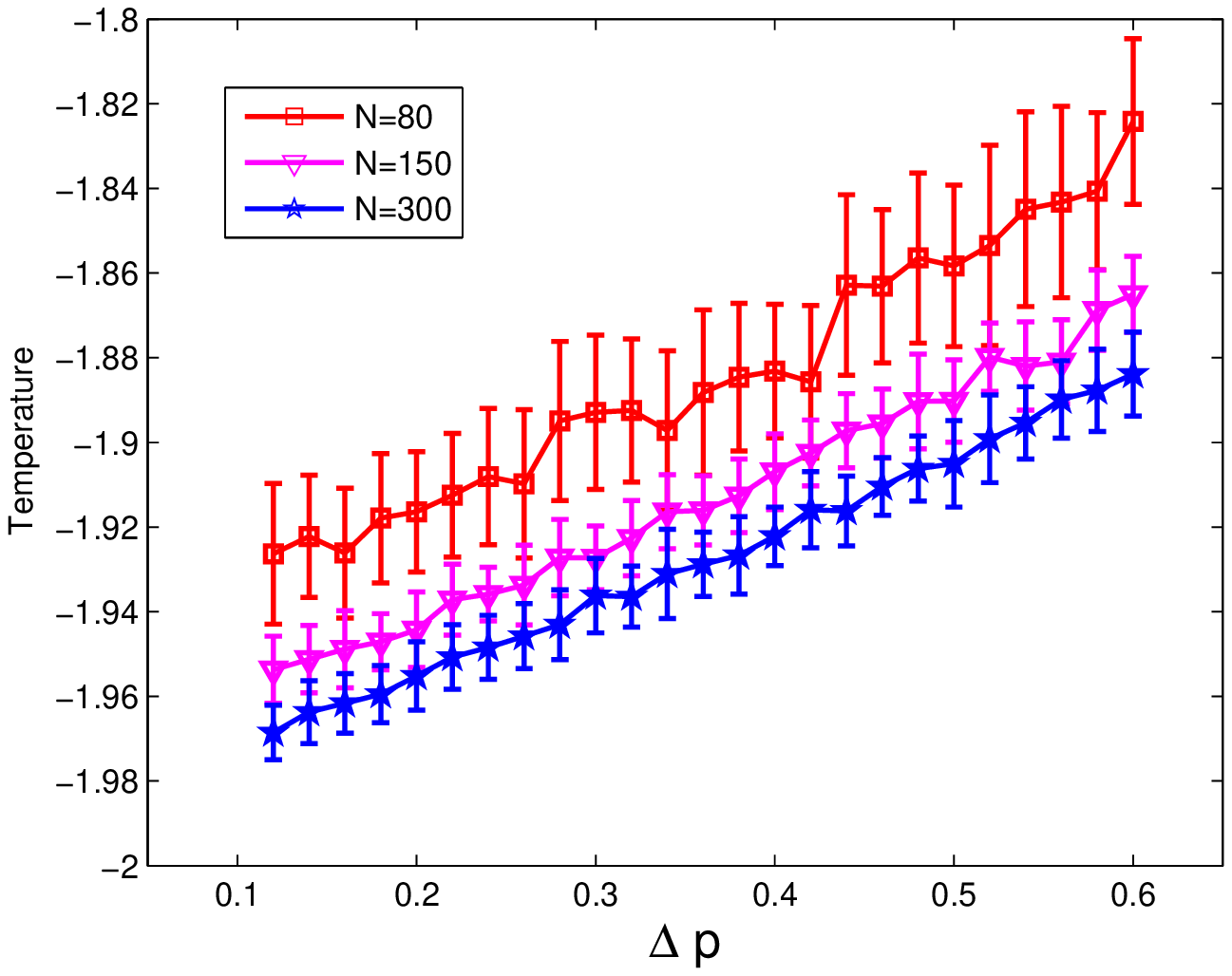}}
 \subfigure[]{\includegraphics[scale=.45]{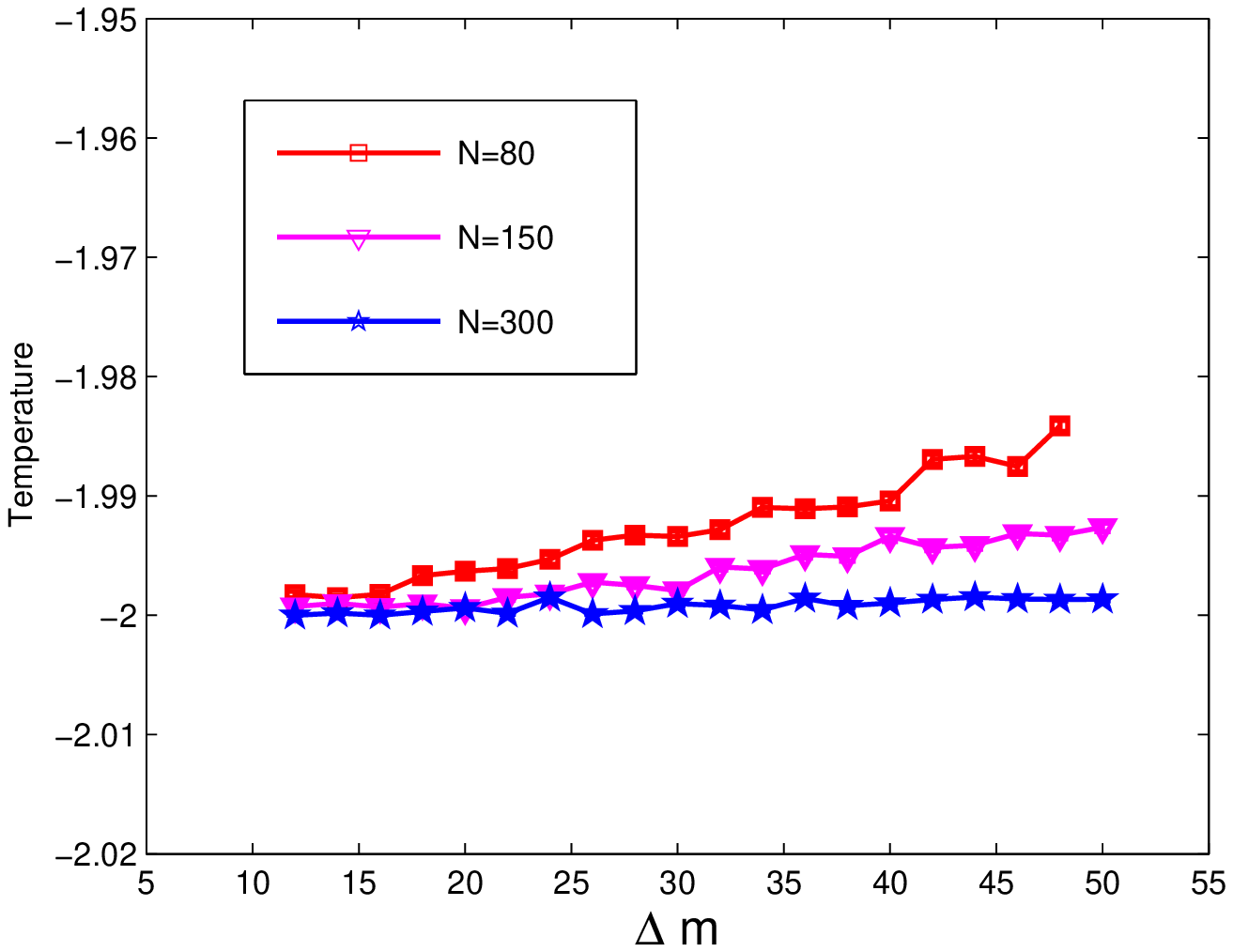}}
 \caption{(Color online) Mean and standard deviation of the temperature versus $\Delta p$ and $\Delta m$ for random graphs with different graph sizes. Red squares: 80 nodes; magenta triangles: 150 nodes; blue stars: 300 nodes.}
\end{figure}

Here we make use of a similar representation of the financial database. Specifically, we employ the correlation-based network to represent the structure of the stock market since many meaningful economic insights can be extracted from the stock correlation matrices \cite{Battiston_2013}\cite{Bonanno_2004}\cite{Caldarelli_2004}. Particularly, to construct the dynamic network,  347 stocks that have historical data from January 1986 to February 2011 are selected \cite{Pero_2011}\cite{Silva_2015}. Then, we use   a time window of 28 days and move this window along time  to obtain a sequence  (from day 29 to day 6004) in which each temporal  window contains a time-series  of the   daily return stock values over a 28-day period. We represent trades between different stocks as a network. For each time window, we compute the cross correlation coefficients between the  time-series for each pair of   stocks, and create connections between them if  the maximum absolute value of the correlation  coefficient is among the highest 5\% of the total cross correlation coefficients. This yields a time-varying stock market network with a fixed number of 347 nodes and varying edge structure for each of 5976 trading days.

\begin{figure}[h]
\centering
 \includegraphics[scale=.5]{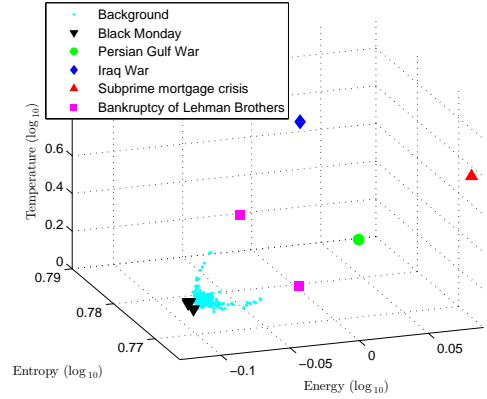}
 \caption{(Color online) The 3D scatter plot of the dynamic  stock correlation network in the thermodynamic space spanned by temperature, average energy and entropy. Cyan dots: Background; black downward-pointing triangles: Black Monday; green circles: Persian Gulf War; blue diamonds: Iraq War; red upward-pointing triangles: Subprime mortgage crisis; magenta squares: Bankruptcy of Lehman Brothers.}
\end{figure}

{\bf Dataset 2:} Is extracted from DNA microarrays that contain the transcriptional profiles for nearly one-third of all predicted fruit fly (Drosophila melanogaster) genes through the complete life cycle, from fertilization to adult. The data is sampled at 66 sequential developmental time points. The fruit fly life cycle is divided into four stages, namely the embryonic (samples 1-30), larval (samples 31-40) and  pupal (samples 41-58) periods together with the first 30 days of adulthood (samples 59-66). Early embryos are sampled hourly and adults are sampled at multiday intervals according to the speed of the morphological changes. At each time point, by comparing each experimental sample to a reference pooled mRNA sample, the relative abundance of each transcript can be measured, which can further be used as a gene's expression level \cite{Arbe_2002}. To represent this gene expression measurements data using a time-evolving network, the following steps are followed \cite{Song_2009}. At each developmental point the 588 genes that are known to play an important  role in the development of the Drosophila are selected. These genes are the nodes of the network. The edges are established based on the distribution of the gene expression values, which can be modeled as a binary pair-wise Markov Random Field (MRF) whose parameter  indicates the strength of undirected interactions between two genes. In other words, two genes are connected when their model parameter exceeds a threshold. This dataset thus yields  a time-evolving Drosophila gene-regulatory network with a fixed number of 588 nodes, sampled at  66 developmental time points.

\begin{figure*}
\centering
 \subfigure[]{\includegraphics[height=2.0in,width=7.2in]{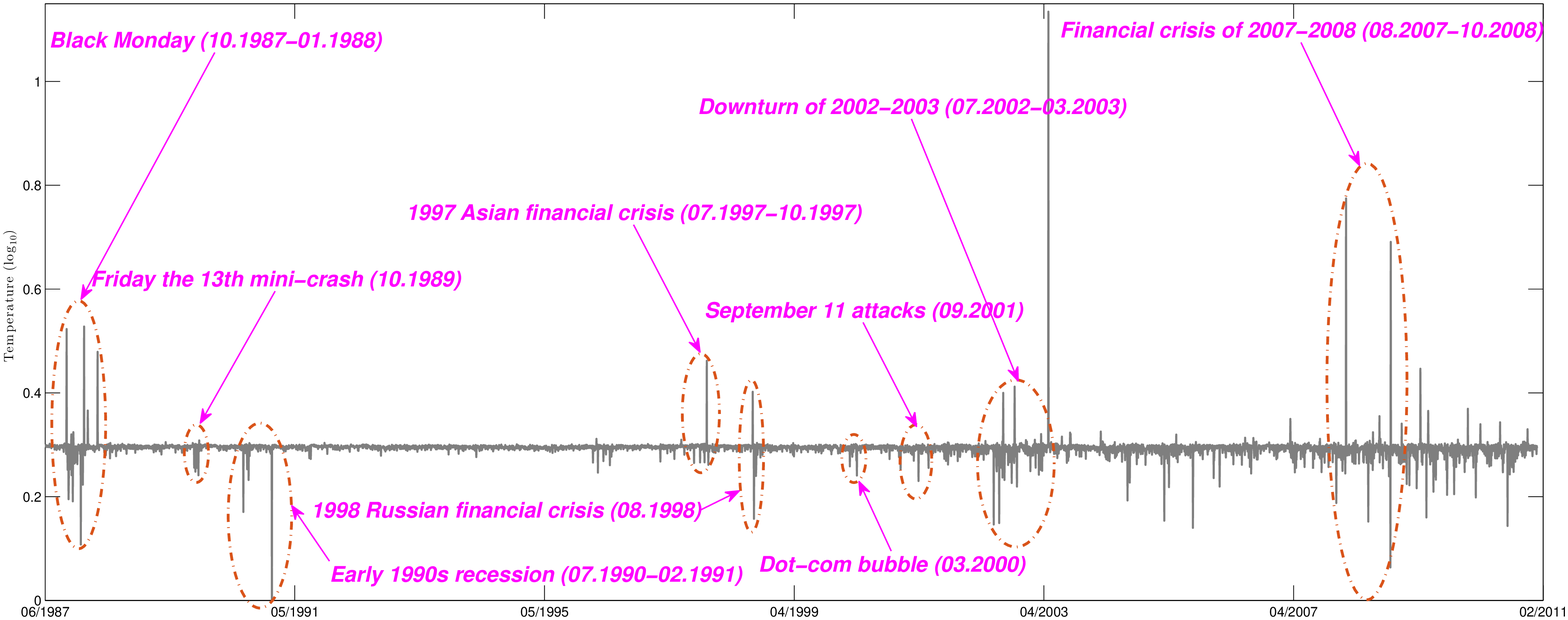}}
 \subfigure[]{\includegraphics[height=2.0in,width=7.2in]{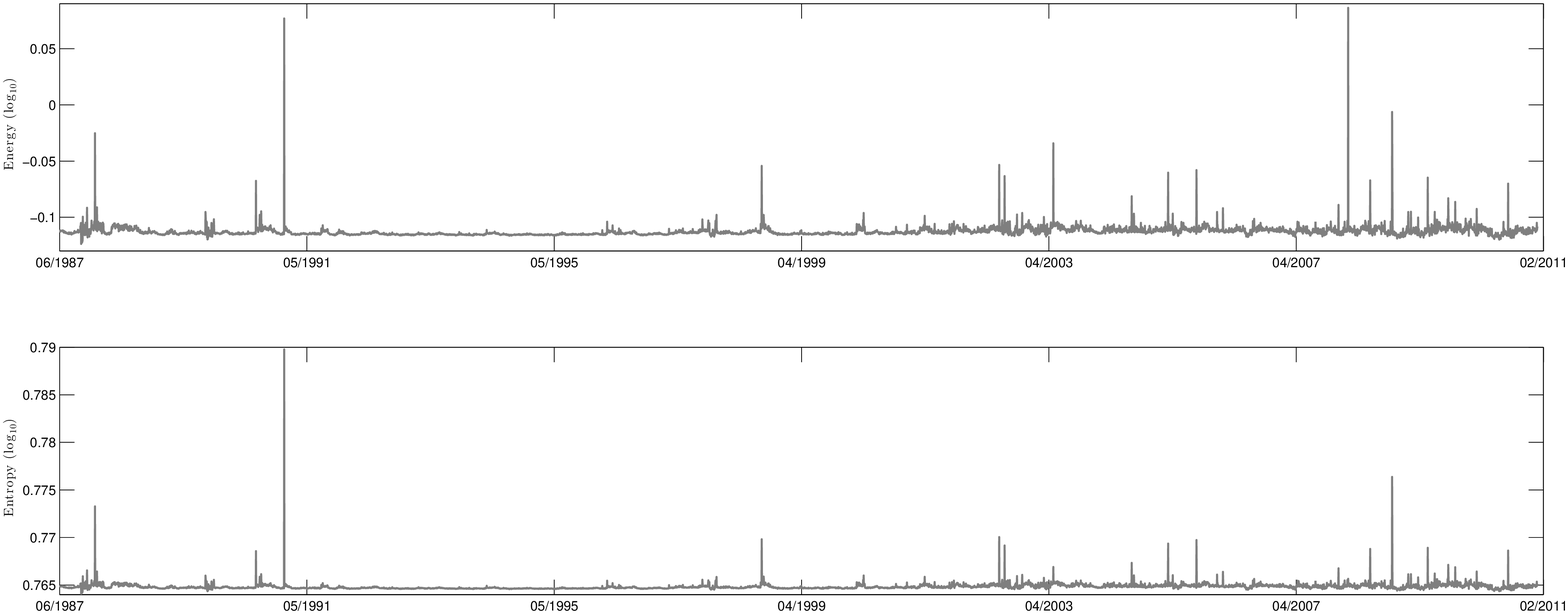}}
 \caption{(Color online) The temperature, average energy and thermodynamic entropy versus time for the dynamic stock correlation network. The known financial crisis periods are identified by ellipses.}
\end{figure*}

\begin{figure*}
\centering
 \subfigure[]{\includegraphics[height=1.8in,width=7.2in]{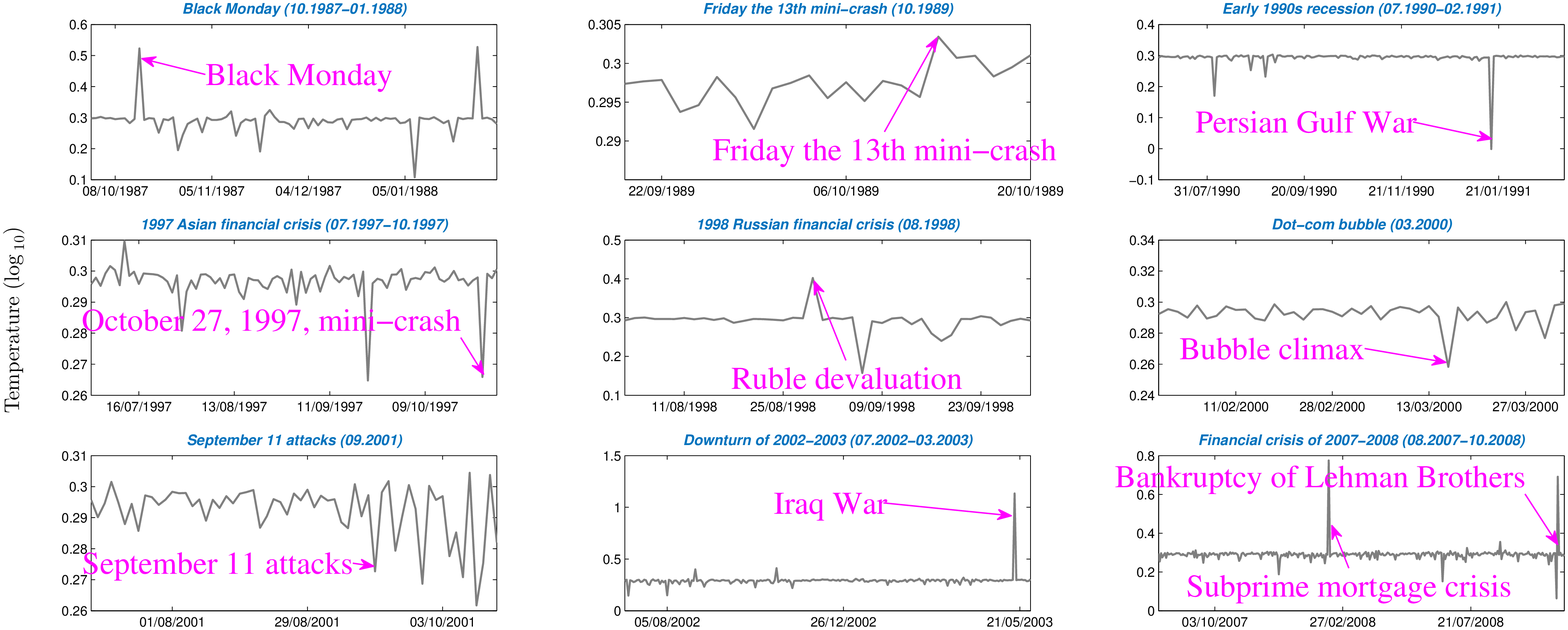}}
 \subfigure[]{\includegraphics[height=1.8in,width=7.2in]{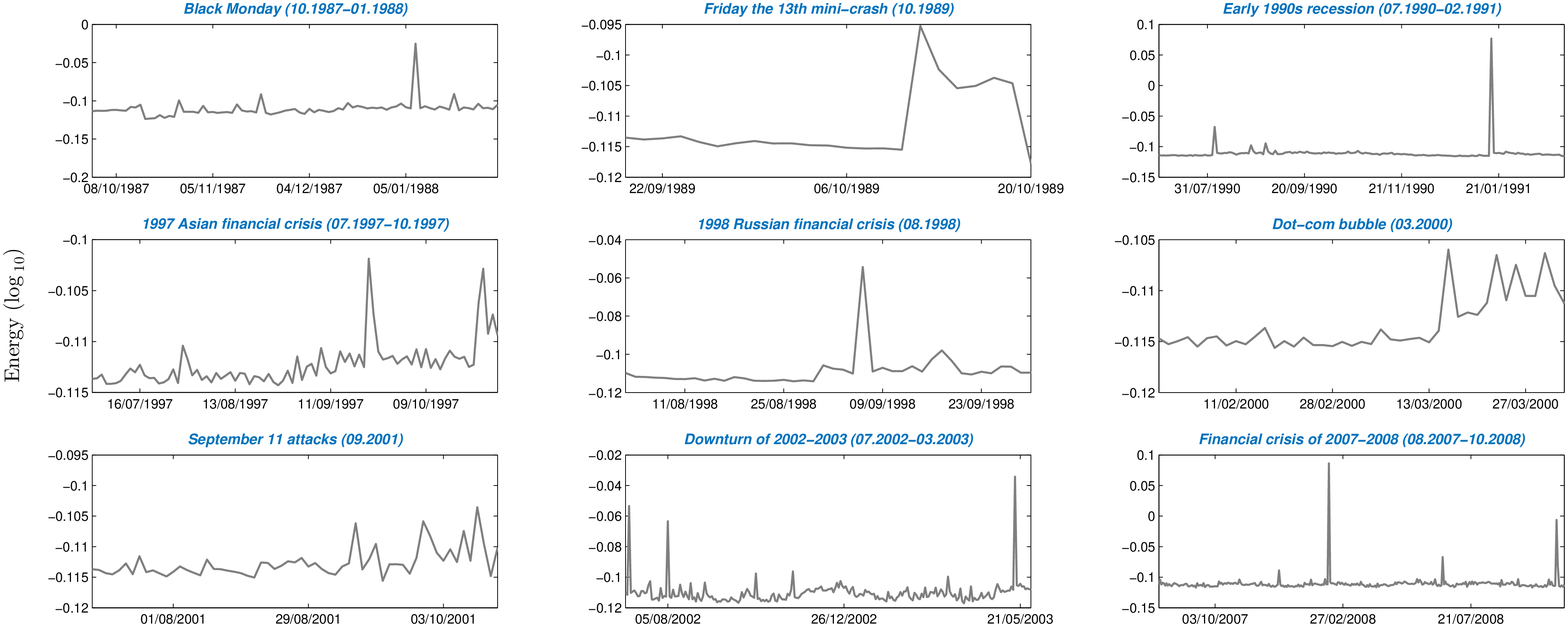}}
 \subfigure[]{\includegraphics[height=1.8in,width=7.2in]{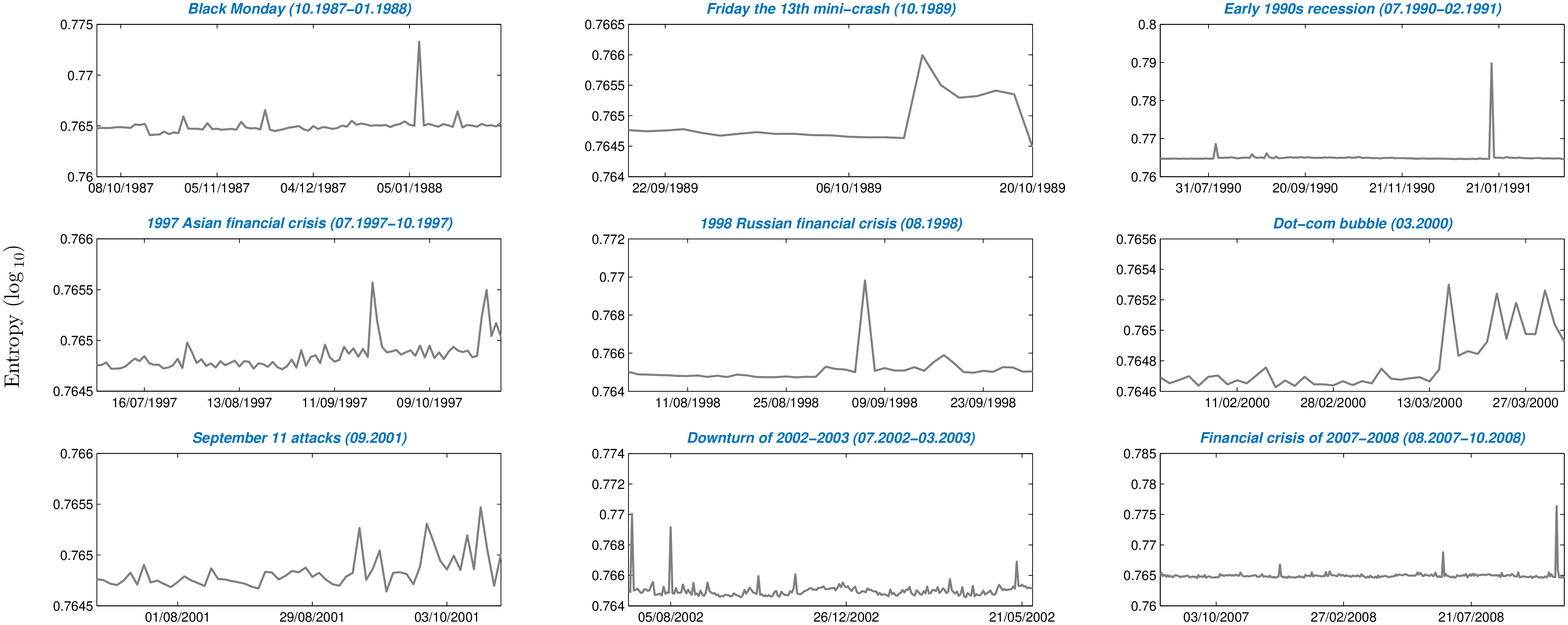}}
 \caption{(Color online) The individual time-series of stock correlation network temperature, energy and entropy for nine different global events that have been identified in Fig. 4.}
\end{figure*}

\subsection{Partition Function and Characteristic Polynomial Approximation}

We commence by examining whether the network Boltzmann  partition function given in Eq.(\ref{partition function}) is well approximated by  the normalized Laplacian quasi characteristic polynomial Eq.(\ref{modified characteristic polynomial}),  as expected from  Eq.(\ref{approximation}). To this end, we first create a large number of random graphs distributed  according to two different models, namely a) the classical Erd\H{o}s-R\'{e}nyi model \cite{Erdos_1959} and b) the Barab\'{a}si-Albert model \cite{Bara_1999}. We randomly generate  500 graphs for each of the two   models using a variety of model parameters. For instance, for the Erd\H{o}s-R\'{e}nyi model, the graph size is between 30 and 1000 and  the connection probability  is $p\in [0.1,0.9]$; for  the Barab\'{a}si-Albert model, the graph size has the same range and the average node degree is bounded between 1 and 20. Then, for each random graph, we compute both the partition function $Z(\beta)$ and the quasi characteristic polynomial $-\ln R(\beta)+N$ for three different values of $\beta$.
The result is shown  as the  scatter plot  in Fig. 1.

\begin{figure}[ht]
\centering
 \subfigure[]{\includegraphics[scale=.45]{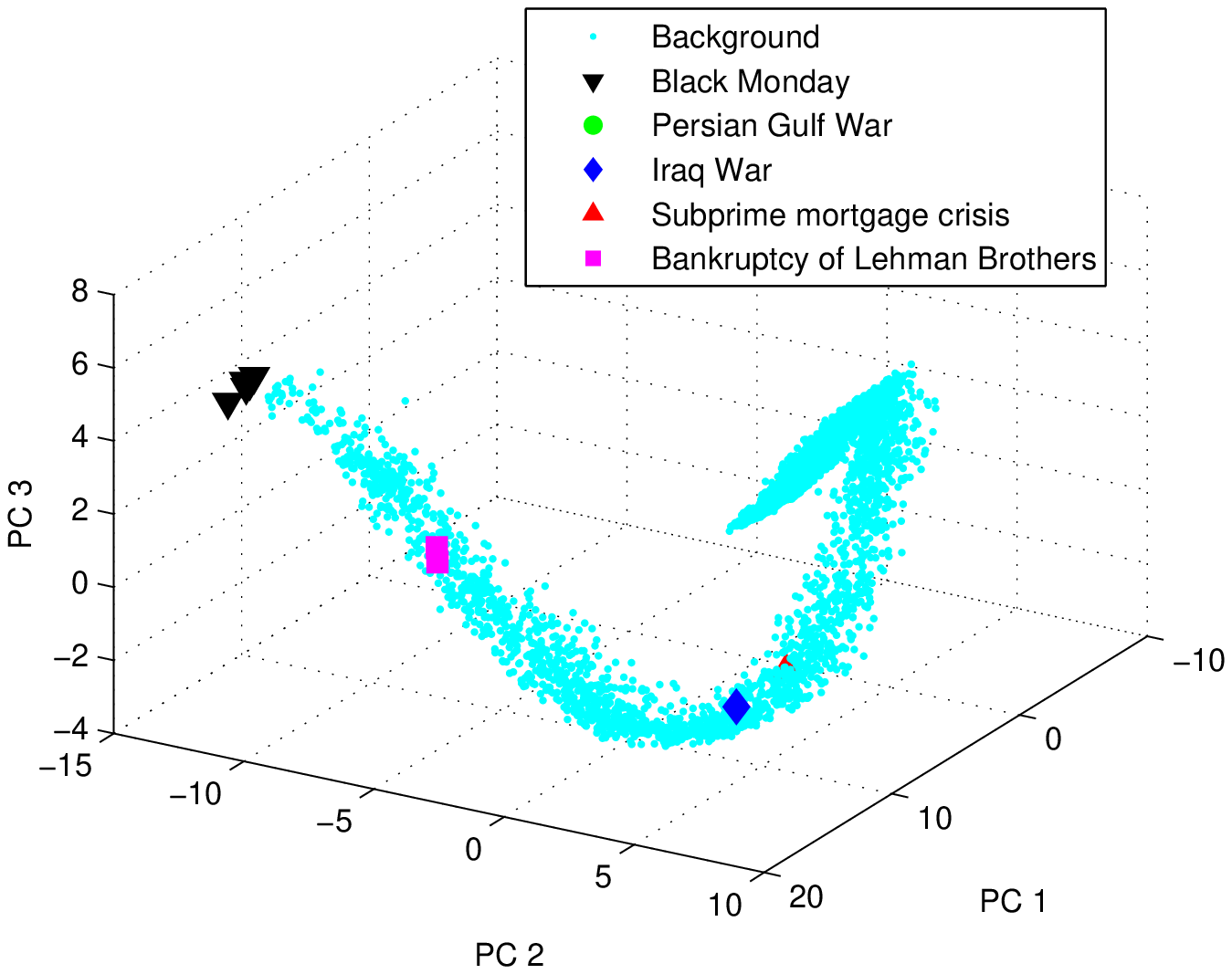}}
 \subfigure[]{\includegraphics[scale=.45]{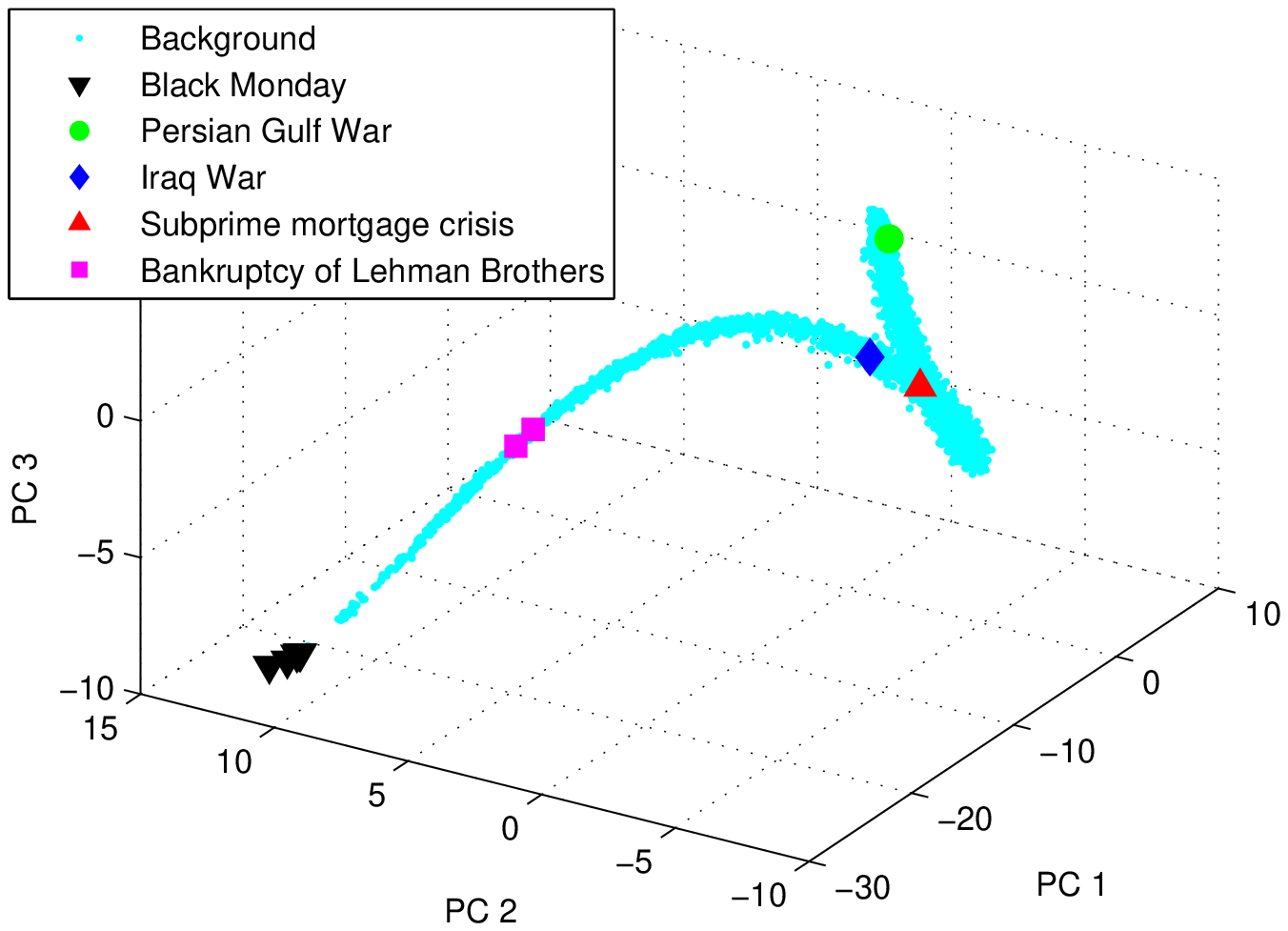}}
 \caption{(Color online) The PCA plots of the dynamic  stock correlation network characterization delivered by different signature methods. Top panel: heat kernel signature; bottom panel: wave kernel signature.}
\end{figure}

\begin{figure}[ht]
\centering
 \subfigure[]{\includegraphics[scale=.45]{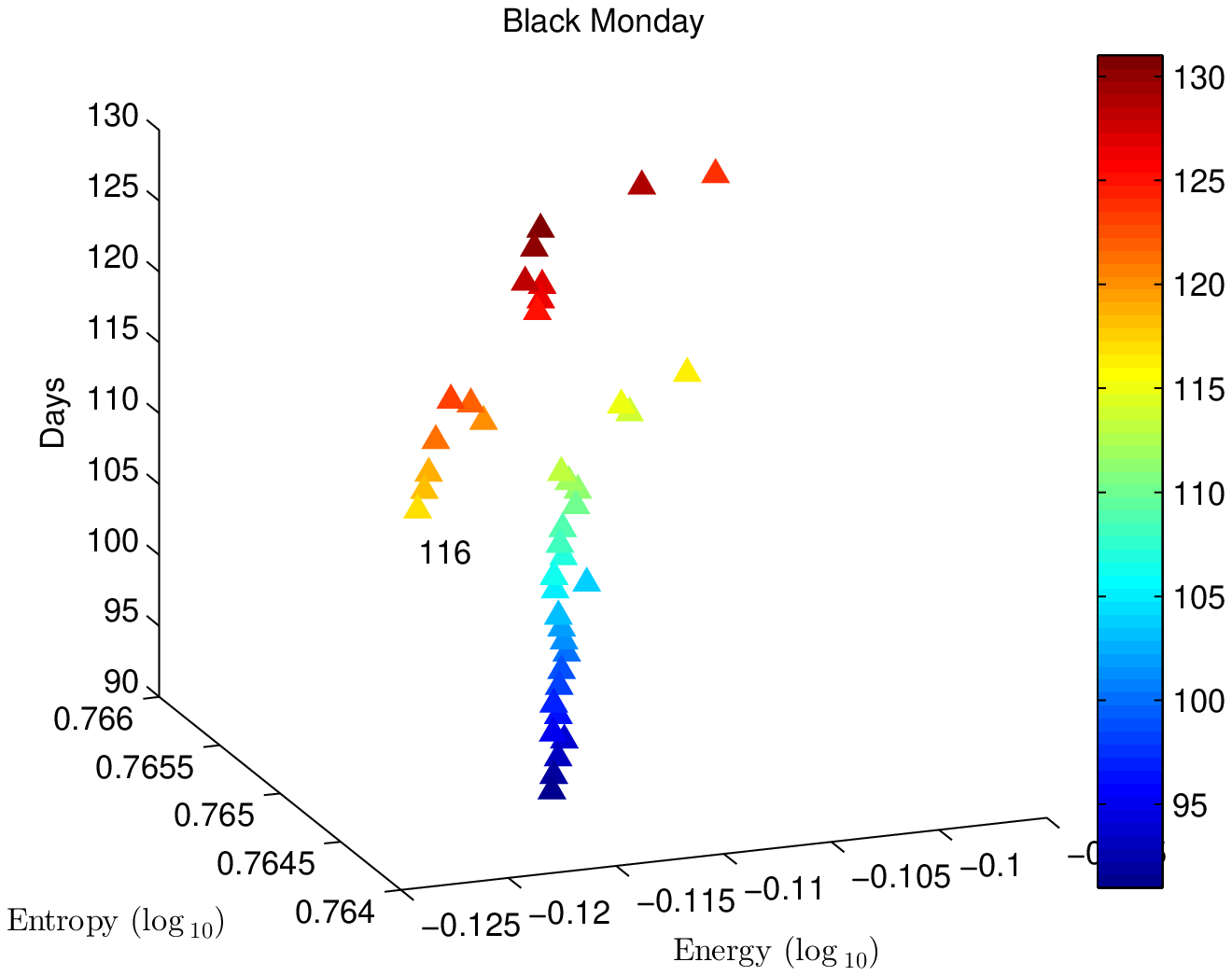}}
 \subfigure[]{\includegraphics[scale=.45]{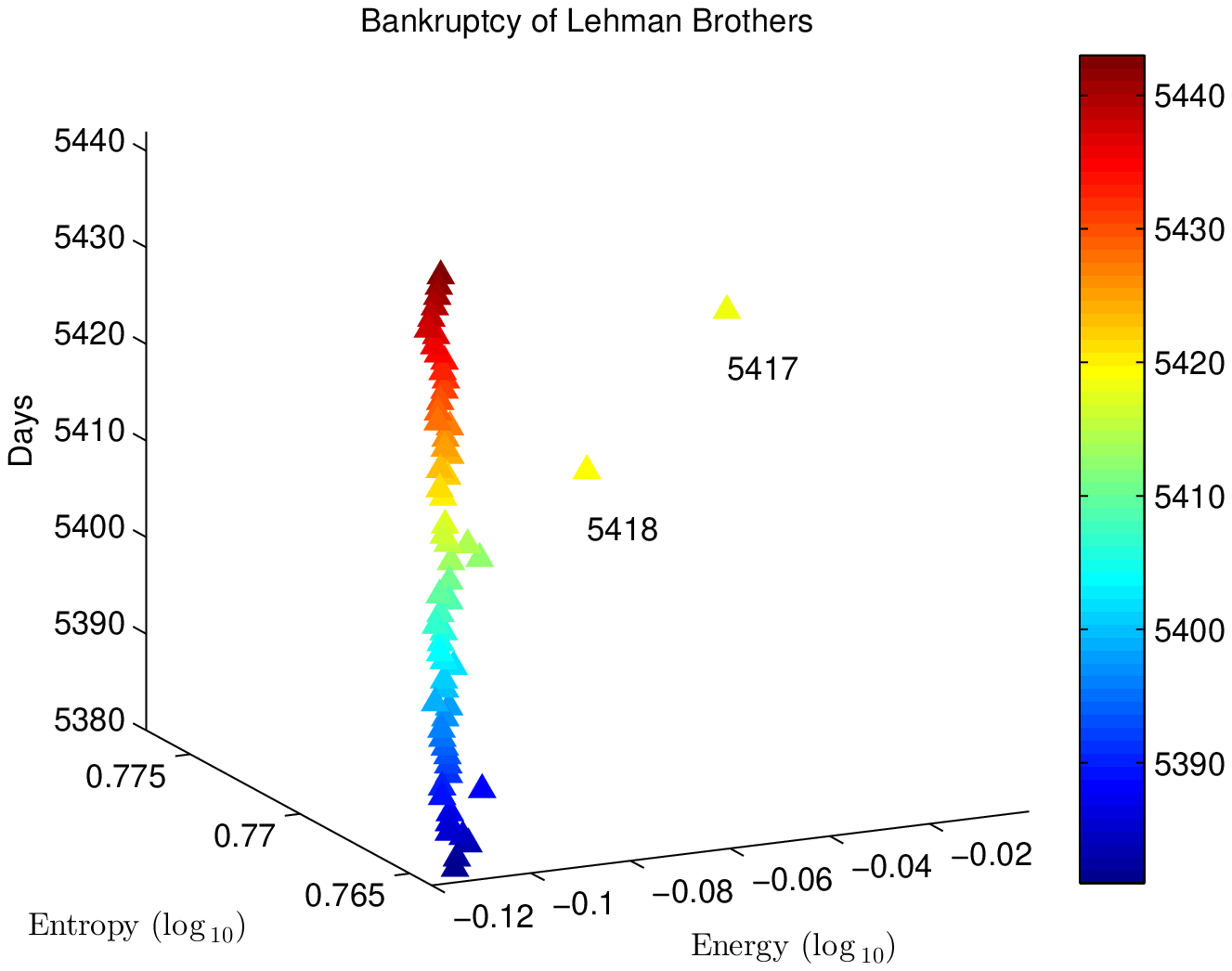}}
 \caption{(Color online) Path of the time-evolving stock correlation network in the entropy-energy-time space during different financial crises. Top panel: Black Monday; bottom panel: Bankruptcy of Lehman Brothers. The colour bar beside each plot represents the date in the time-series.}
\end{figure}

The most striking feature in this figure is that although $\beta$ takes on different values, the vast majority of the corresponding data points are close to the diagonal line $y=x$. This result empirically proves that the partition function $Z(\beta)$ is always very accurately approximated by the characteristic polynomial $-\ln R(\beta)+N$ for different types of random graphs, as shown in Eq.(\ref{approximation}).

\subsection{Temperature and Network Structure}

In this subsection, we investigate the relationship between the thermodynamic variables developed and the structural change of networks. Specifically, we aim at exploring how the temperature fluctuates when a graph experiences various degrees of evolutionary change. To this end, we commence by constructing a complete graph with 80 nodes, and randomly deleting its edges with a probability $p \in [0,0.2]$. Then, we start from the same complete graph, and with probability $p+\Delta p$, we again delete edges in the graph randomly. Using these two random graphs, we compute the temperature according to Eq.(\ref{temperature}). We repeat the process for different values of $\Delta p \in [0.1,0.6]$ (100 realizations each), which indicate the different degrees of structural change during graph evolution. We then repeat the analysis for graphs with 150 nodes and 300 nodes respectively and produce a plot showing the mean and standard deviation (shown as error bar)  of the temperature against $\Delta p$ for a large number of random graphs with different sizes.

The most important feature in Fig. 2(a) is that as $\Delta p$ increases, the mean values of the temperature for all three graph sizes grow. Moreover, the variance of temperature also increases gradually with the increase of $\Delta p$. This is because the variance of the ratio $(Q_1-Q_2)/(J_1-J_2)$ becomes large when there is a dramatic structural change in the time-evolving network, resulting in  a significant change of the value of temperature. Moreover, when $\Delta p$ remains  small, the temperature remains relatively stable. This result agrees  with expression for temperature in Eq.(\ref{temperature}).  Slight evolutionary changes lead to a small value of  $(Q_1-Q_2)/(J_1-J_2)$, the value of temperature  then stabilizes at $-2$.

In order to demonstrate that fluctuations in temperature are caused by  structural changes in the arrangement of edges in a network, rather than by difference in edge number between two networks, we provide the following empirical analysis. We first create a regular graph of 80 nodes with degree $m=10$, and create a second  regular graph that has the same graph size, but with a greater degree $m+\Delta m$. Thus, the temperature due to fluctuations  between these two networks can be computed. For each $\Delta m=12,14,\ldots,50$, we again produce 100 realizations of the  graphs. We then plot the mean and standard deviation of temperature against $\Delta m$ for different graph sizes in Fig. 2(b). For random graphs with various node number, although there are some fluctuations, the temperature is almost constant   despite the fact that  the degree difference varies significantly. This is because there is no significant change in the  internal structure of the network during such an evolution. This result confirms that the thermodynamic characterizations are effective  in capturing the changes in internal  structure  of time  evolving networks.

The value of the temperature deserves further comment. In this experiment $T$ is always negative, this is because the first term in the temperature expression Eq.(\ref{temperature}) has a minus sign. It is worth stressing that this sign appears naturally from the temperature development  and  it does not mean the temperature is negative physically.

\subsection{Thermodynamic Measures for Analyzing Network Evolution}

We explore whether the thermodynamic measures can be used as an effective tool for better understanding the evolution of realistic complex networks. To commence, we explore the evolutionary behaviour of the NYSE stock market by applying our thermodynamic characterization method to the dynamic networks in Dataset 1. At each time step, we compute the average energy, entropy and temperature according to Eq.(\ref{average energy}), Eq.(\ref{entropy}), and Eq.(\ref{temperature}) respectively. This allows us to investigate how these network thermodynamic variables  evolve with time and whether critical  events can be detected in the network evolution.

Figure 3 is  a 3-dimensional scatter plot showing the thermodynamic variables  for the time-evolving stock correlation network. It represents a  thermodynamic space spanned by average energy, entropy and temperature. The thermodynamic distribution of networks clearly shows a strong manifold structure. The outliers, on the other hand, indicate  singular  global events. Examples include  Black Monday (black downward-pointing triangles) \cite{Brow_2007}, the Persian Gulf War and Iraq War (green circles and blue diamonds respectively), and the subprime mortgage crisis (red upward-pointing triangles) together with the bankruptcy of Lehman Brothers (magenta squares).

The individual time-series for different thermodynamic variables, i.e., temperature, energy and entropy are shown in Fig. 4. There are a number of important observations. First, most of the significant fluctuations  in the individual time-series of thermodynamic variables successfully correspond to some realistic serious financial crises, e.g., Black Monday \cite{Brow_2007}, Friday the 13th mini-crash \cite{Jenk_2002}, September 11 attacks and the bankruptcy of Lehman Brothers \cite{Moll_2008}. The reason for this is  that the stock market network experiences dramatic structural changes when a financial crisis occurs. For instance, during the dot-com bubble period \cite{Ande_2010}, a significant  number of Internet-based companies were founded, leading to a rapid increase of both stock prices and market confidence.  This considerably modified both  the inter-relationships between stocks and the resulting structure of the entire market, which can be captured by the thermodynamic characterization. Another interesting feature in the figure is that  the stock correlation network structure becomes considerably unstable after entering the 21st century, compared to that before year 2000. Particularly, there are a great number of significant fluctuations in all three time-series in recent years, which is due to the outbreak of the global recession and financial crisis that began in 2007.

\begin{figure}[h]
\centering
 \includegraphics[scale=.5]{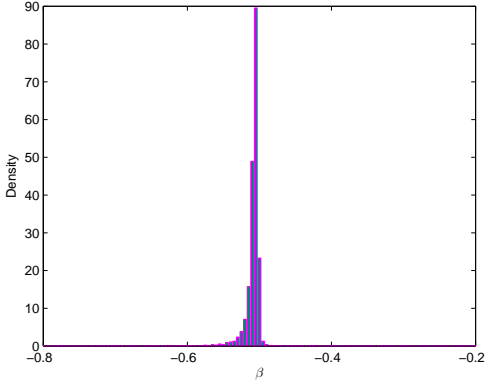}
 \caption{(Color online) The normalized histogram of $\beta$, defined as $\beta=1/kT$, for the dynamic stock correlation network.}
\end{figure}

\begin{figure}[ht]
\centering
 \includegraphics[scale=.5]{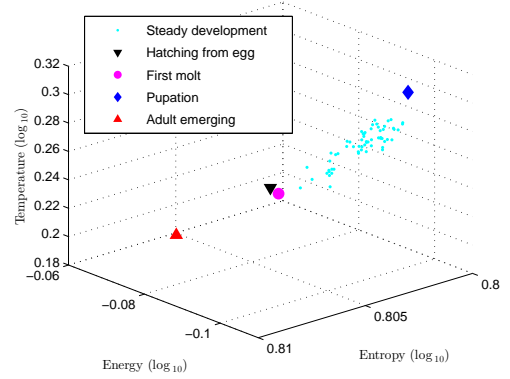}
 \caption{(Color online) The 3D scatter plot of the dynamic Drosophila gene regulatory network in the thermodynamic space spanned by temperature, average energy and entropy. Cyan dots: Steady development; black downward-pointing triangle: Hatching from egg; magenta circles: First molt; blue diamonds: Pupation; red upward-pointing triangles: Adult emerging.}
\end{figure}

To see more clearly the detail of how the thermodynamic variables change over time during the different financial crises, in Fig. 5 we show all three thermodynamic variable time-series for the nine global events identified in Fig. 4. From Fig. 5(a), the most striking observation is that almost all of the largest peaks and troughs can find their realistic financial crisis correspondences, which shows the thermodynamic characterization is sensitive to network structural changes. Also, different global events exhibit different detailed behaviours. For example, both wars (Persian Gulf and Iraq) dramatically change the network structure in a short time, which are shown as a sharp trough and peak in the corresponding time-series. Moreover, the September 11 attacks clearly have a persistent influence on the stock market since the network temperature fluctuates significantly after the attack. The reason for this is that different financial crises affect the stock network structure in different ways. Specifically, some crises lead the degree-products for both triangles $Q$ and edges $J$ increase or decrease simultaneously (Black Monday, Iraq War, the subprime mortgage crisis, etc.), and as a result $(Q_1-Q_2)/(J_1-J_2)$ is positive and the temperature increases. In contrast, some events lead to the result that $J$ and $Q$ change in a different direction, which means that $(Q_1-Q_2)/(J_1-J_2)$ is negative and the temperature decreases accordingly, such as Persian Gulf War, the mini−crash on October 27, 1997 and the dot-com bubble climax.

We now compare our thermodynamic representation with a number of methods from the  spectral  analysis of graphs, namely the heat kernel signature \cite{Sun_2009} and the wave kernel signature \cite{Aubry_2011}. Figure 6 shows 3-dimensional scatter plots obtained from the principal component analysis (PCA) of network  characterizations delivered by these two methods respectively. Both plots show a compact manifold structure. However, only the Black Monday (black triangles) can be identified. The critical points representing other financial events such as  the subprime mortgage crisis and the bankruptcy of Lehman Brothers, do not deviate from the manifold structure, which means that these events cannot be detected. This illustrates that the thermodynamic characterization provides an effective method for analyzing financial network evolution, which smooths the manifold structure  while preserving information concerning  significant changes in network structure.

\begin{figure*}
\centering
 \includegraphics[height=2.8in,width=7.2in]{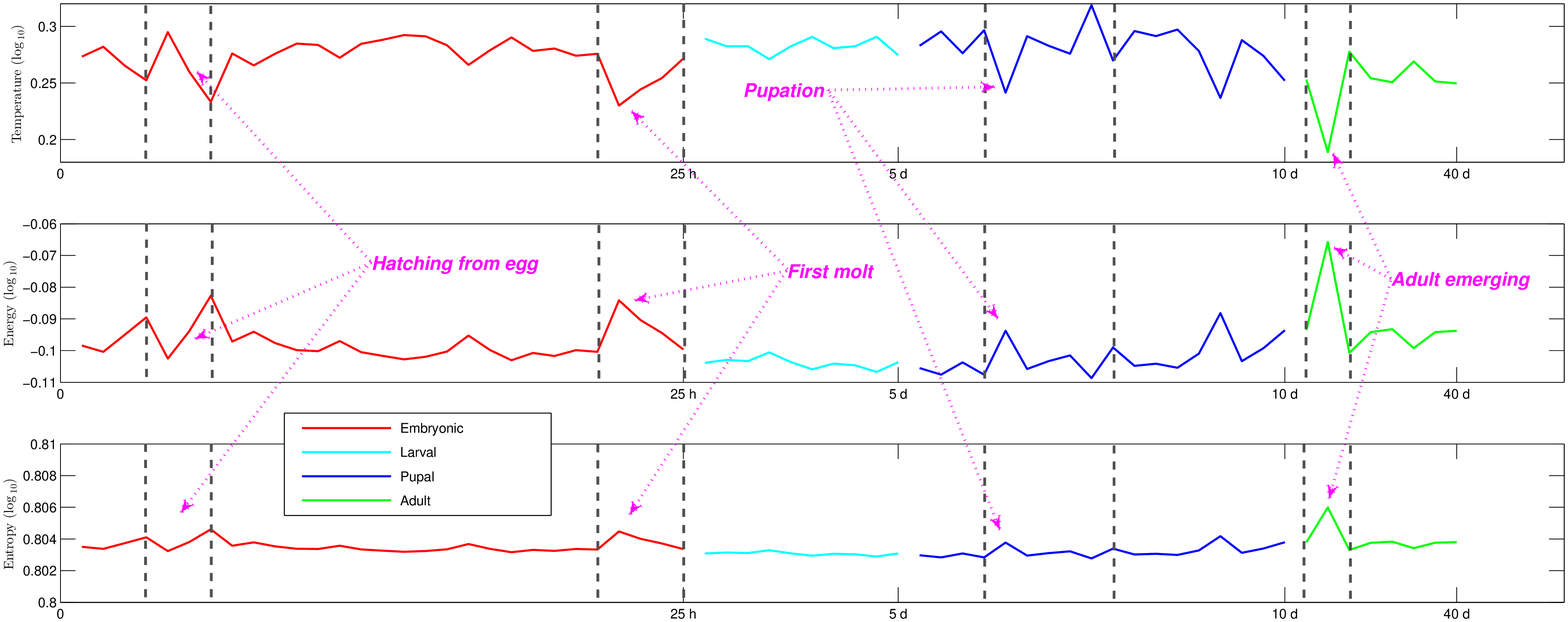}
 \caption{(Color online) The temperature, average energy and the thermodynamic entropy versus time for the dynamic Drosophila gene regulatory network. The important morphological changes are identified by arrows. Red line: Embryonic; cyan line: Larval; blue line: Pupal: green line: Adulthood.}
\end{figure*}

\begin{figure}[ht]
\centering
 \subfigure[]{\includegraphics[scale=.45]{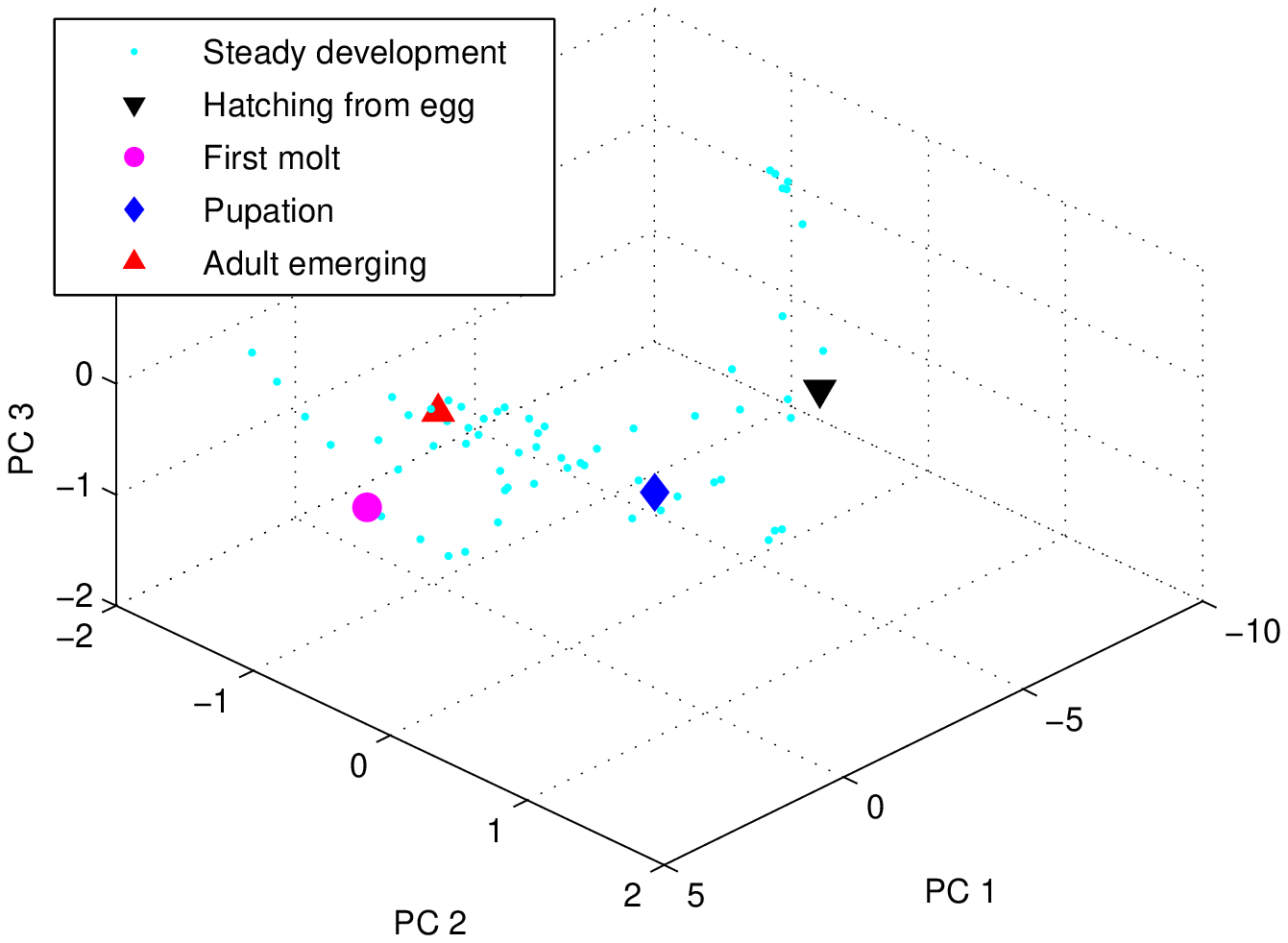}}
 \subfigure[]{\includegraphics[scale=.45]{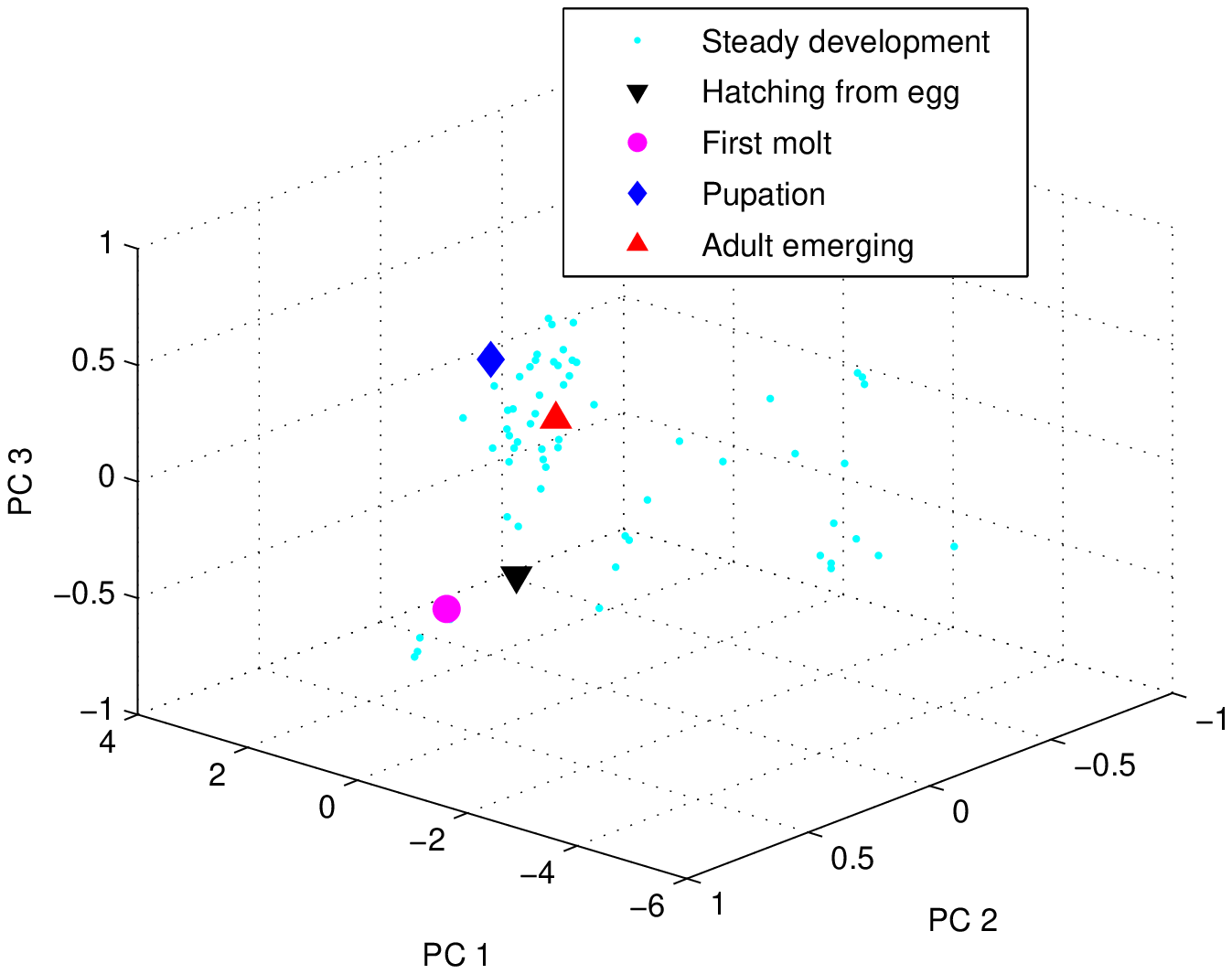}}
 \caption{(Color online) The PCA plots of the dynamic Drosophila gene regulatory network characterization delivered by different signature methods. Top panel: heat kernel signature; bottom panel: wave kernel signature.}
\end{figure}

We now focus on   two different   financial crises in more detail,  and explore how the stock market network structure changes with time according to the  thermodynamic variables. In Fig. 7 we show a set of points indicating  the path of  the stock network in the entropy-energy space  with time during a) Black Monday and  b) the Lehman Brothers bankruptcy.  The colour bar beside each plot represents the date in the time-series. The top panel shows that before Black Monday (blue and green triangles), the network structure remains  stable. Neither  the network entropy nor the  average energy  change significantly. However, during Black Monday (from day 116), the network undergoes a considerable change in structure since the entropy decreases dramatically and energy increases significantly. After the crisis, the stock correlation network gradually returns to its normal state. A different behaviour can be observed  concerning the Lehman Brothers bankruptcy which is shown in the bottom panel. The stock network undergoes a significant crash in which the  network structure undergoes a significant change, as signalled by a large increase in both network energy and entropy. More importantly, the crash is followed by a quick recovery. Hence, our thermodynamic  representation can be used to both characterize and distinguish between different financial crises.

In Fig. 8 we provide a normalized histogram of $\beta$ for this time-evolving stock correlation network. The most striking  feature is that the vast majority of this parameter is between -0.6 and -0.4. This result shows empirically, that for real-world complex networks, the approximation between the Boltzmann partition function and the quasi characteristic polynomial of normalized Laplacian matrix Eq.(\ref{approximation}) is valid.

We now turn our attention to the fruit fly network, i.e., the Drosophila gene regulatory network contained in Dataset 2. In Fig. 9, we again show a 3-dimensional scatter plot of the time-varying  thermodynamic variable space. Unlike the NYSE data for the stock, here the data points do not display a clear manifold in the thermodynamic space. This is because there are only 66 time epochs in the time-series of the gene regulatory network. Nevertheless, some critical morphological changes can still  be identified, such as the egg hatching (black triangle), molt (magenta circle) and pupation (blue diamond). More importantly, the red triangle, representing the most significant morphological change, namely the emergence of the adult is separated by the greatest distance from the remainder of the developmental samples. This indicates that the thermodynamic characterization successfully captures the evolutionary changes in the underlying dynamic network.

Figure 10 shows the separate time-series of temperature, energy and  entropy for the fruit fly network. Also shown in this figure are a number of critical evolutionary events, which are indicated by arrows and four developmental stages, which are  shown in different colours. In  the plot, the early development of embryo, which is represented using the red line (embryonic period) shows some significant fluctuations. This is attributable to strong and rapidly changing gene interactions, because of the need for rapid organism  development. Moreover, in the pupal stage, there are also considerable fluctuations. This is attributable to the fact that during this period, the pupa undergoes a number of significant pupal-adult transformations. As the organism evolves into an adult, the gene interactions which control its growth begin to slow down. Hence the green line (adulthood) remains stable (after the adult emerges).

We again provide a comparison between our thermodynamic representation and the heat kernel signature together with the wave kernel signature analyses on this biological data. To this end, we apply the principal component analysis (PCA) to the network  characterizations delivered by these two methods and obtain the 3-dimensional scatter plots in Fig. 11. Comparing to Fig. 9, it is difficult to distinguish the time points when significant morphological changes take place between those representing steady evolutionary development. This observation confirms that the thermodynamic characterization is not only effective in the financial domain, but also provides some useful insights to analyze the biological data.

Finally, in Fig. 12 we show a normalized histogram of $\beta$ for the Drosophila gene regulatory network. The main conclusion from the plot is that result  $\beta$  is most densely distributed over the  interval $(-0.6,-0.45)$, empirically showing that $|\beta|$ takes on a small value such that $r(\beta)=o[\ln R(\beta)]$, which again confirms the validity of the approximation obtained in Eq.(\ref{approximation}).

\begin{figure}[ht]
\centering
 \includegraphics[scale=.5]{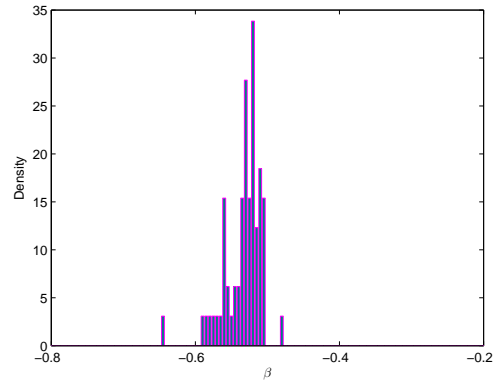}
 \caption{(Color online) The normalized histogram of $\beta$, defined as $\beta=1/kT$, for the dynamic Drosophila gene regulatory network.}
\end{figure}

\section{Conclusions}

In this paper, we show how a characteristic polynomial can be used to approximate the Boltzmann partition function of a network. We commence from a quasi characteristic polynomial computed from the normalized Laplacian matrix of a graph and show how this polynomial is linked to the Boltzmann partition function of the graph, when the graph Hamiltonian is defined by the normalized Laplacian operator. This allows us
to derive  a thermodynamic representation of  network structure which can be used to visualize and understand the evolution of time-varying networks. Under the assumption that the network is of constant volume,
we provide  approximate  expressions for a number of thermodynamic network variables,  including the entropy, average energy and temperature.

We  evaluate the method experimentally using data representing  a variety of real-world  complex systems, taken from the  financial  and biological domains. The experimental results demonstrate that the thermodynamic variables are efficient in analyzing the evolutionary properties of dynamic networks, including the detection of abrupt changes and phase transitions in structure or other distinctive periods in the evolution of time-varying complex networks.

The method does though appear to have some limitations. For instance it does appear sensitive to random fluctuations in network structure, not associated with identifiable events in the time-series studied. Also
critical events do not necessarily give rise to unique patterns.

In the future, it would be interesting to see what features the thermodynamic network variables reveal in additional domains, such as human functional magnetic resonance imaging data. Another interesting line of investigation would be to explore if the thermodynamic framework can be extended to the domains of dynamic directed networks, edge-weighted networks, labelled networks and hypergraphs. Finally, it would be intriguing to investigate whether partition functions from different quantum statistics, such as Bose-Einstein partition function and Fermi-Dirac partition function, together with Ihara zeta function can be applied to network science to provide a way to probe larger structures.

\begin{acknowledgments}

Cheng Ye is supported by the National Natural Science Foundation of China (Grant No.61503422). F. N. Silva acknowledges CAPES. C. H. Comin thanks FAPESP (Grant No. 11/22639-8) for financial support. T. K. DM. Peron acknowledges FAPESP (Grant No. 2012/22160-7) for support. F. A. Rodrigues acknowledges CNPq (Grant No. 305940/2010-4), FAPESP (Grant No. 2011/50761-2 and 2013/26416-9) and NAP eScience - PRP - USP for financial support. L. da F. Costa thanks CNPq (Grant No. 307333/2013-2) and NAP-PRP-USP for support. This work has been supported also by FAPESP grants 12/50986-7 and 11/50761-2.

\end{acknowledgments}

\bibliography{Bibliography}

\end{document}